\documentstyle[prl,aps,twocolumn,epsf]{revtex}

\def \I{{\rm i}}
\def \D#1{{$D_#1$}}

\begin{document}

\twocolumn[\hsize\textwidth\columnwidth\hsize\csname
@twocolumnfalse\endcsname

\title{Quantum-scissors device for optical state truncation:\\
A proposal for practical realization }

\author{
\c{S}ahin Kaya \"Ozdemir,$^{(1)}$\cite{e-mail}  Adam
Miranowicz,$^{(1,2)}$ Masato Koashi,$^{(1)}$ and Nobuyuki
Imoto$^{(1,3)}$}

\address{
$^{1}$ CREST Research Team for Interacting Carrier Electronics,
School of Advanced Sciences, \\Graduate University for Advanced
Studies (SOKEN), Hayama, Kanagawa 240-0193, Japan\\
$^{2}$ Nonlinear Optics Division, Institute of Physics, Adam
Mickiewicz University, 61-614 Pozna\'n, Poland\\
$^{3}$ NTT Basic Research Laboratories, 3-1 Morinosato Wakamiya,
Atsugi, Kanagawa 243-0198, Japan}

\date{Received 3 July 2001}
\pagestyle{plain} \pagenumbering{arabic} \maketitle

\begin{abstract}
We propose a realizable experimental scheme to prepare
superposition of the vacuum and one-photon states by truncating an
input coherent state. The scheme is based on the quantum scissors
device proposed by Pegg, Phillips, and Barnett {[}Phys. Rev. Lett.
{\bf 81}, 1604 (1998){]} and uses photon-counting detectors, a
single-photon source, and linear optical elements. Realistic
features of the photon counting and single-photon generation are
taken into account and possible error sources are discussed
together with their effect on the fidelity and efficiency of the
truncation process. Wigner function and phase distribution of the
generated states are given and discussed for the evaluation of the
proposed scheme.

\vspace{1mm} DOI: 10.1103/PhysRevA.64.0638XX \hspace{.8cm} PACS
number(s): 42.50.Dv, 42.50.Ar, 03.65.Ta, 03.67.-a \vspace{5mm}
\end{abstract}
]

\section{Introduction}

\begin{figure*}[h]
\vspace*{-3mm}\hspace*{-3cm} \epsfxsize=11cm
\epsfbox{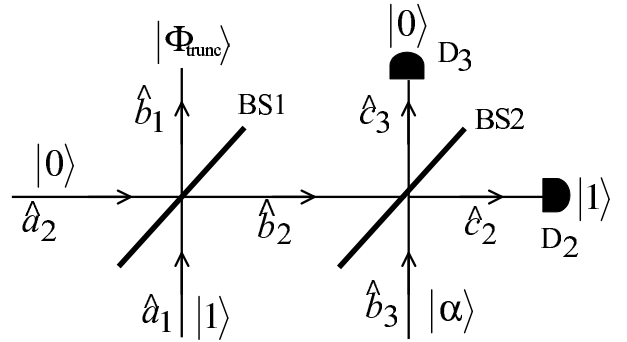}\vspace*{-3.0cm}%
\caption{Schematic configuration of the quantum scissors device
(QSD). BS1, BS2, beam splitters; \D2, \D3, photon counting
detectors; $|\alpha\rangle$, $|0\rangle$, $|1\rangle$, coherent,
vacuum, and single-photon states, respectively; $|\Phi_{\rm
trunc}\rangle$, truncated output state. \label{fig01}}
\end{figure*}

There has been a growing interest in the generation and
engineering of quantum states of light. Over the last decade,
various schemes for preparation of Fock states \cite{Fock} and
their arbitrary finite superpositions
\cite{state1,Peg98,Bar99,Dak99,Dar00,Res01,Kon00,Leo97,state2}
have been developed. The motivation behind these efforts is the
possible applications of nonclassical states of light in quantum
communication and information processing. Such states have been
shown to be generated by nonlinear media or by conditional
measurements at the output ports of beam splitters. For example,
the method proposed by Dakna et al. \cite{Dak99} relies on an
alternate application of coherent displacement and photon adding
(and/or subtracting) via conditional measurements on beam
splitters for the generation of several different types of
nonclassical states. That scheme consists of photon-counting
devices, high-transmittance beam splitters, and the condition of
no-photon detection at the detectors. Another interesting scheme
proposed by D'Ariano et al. \cite{Dar00} is based on ring cavity
and Kerr medium. However, the simplest scheme is the one proposed
by Pegg and co-workers \cite{Peg98,Bar99}. This scheme (see Fig.
\ref{fig01}), referred to as the {\em quantum scissors device}
(QSD), enables generation of the finite superpositions of number
states by truncating a coherent state. Recently, Resch  et al.
proposed and experimentally demonstrated a QSD-like
state-preparation technique based on conditional coherence
\cite{Res01}.

The quantum scissors device exploits three fundamental concepts of
quantum mechanics: (a) Entanglement, mixing of vacuum and a single
photon at the first beam splitter (BS1) creates an entangled state
and opens a quantum channel; (b) measurement, a physical system
can be brought to a desired state by a conditional measurement;
and (c) nonlocality, vacuum and single-photon components of the
coherent state at ${\hat b}_{3}$ are generated at the ${\hat
b}_{1}$ mode without any light going from ${\hat b}_{3}$ of the
second beam splitter (BS2) to ${\hat b}_{1}$ mode of BS1.
Recently, the basic idea of the QSD has been slightly modified to
generate the superposition of vacuum, one-photon, and two-photon
states \cite{Kon00}. An interferometric scheme equivalent to a QSD
with tunable beam splitters has been proposed by Paris to prepare
arbitrary superposition states \cite{Paris00}. It has also been
shown that the basic QSD scheme can be applied as a teleportation
device for superposition states \cite{Vil99}. No proposal has been
made concerning the practical scheme of the QSD, which considers
realistic models for the detectors and sources.

In this paper, our main interest is to propose and study an
experimental QSD scheme for producing superposition of the vacuum
and one-photon states, $C_{0}|0\rangle + C_{1}|1\rangle$, which is
the simplest optical-qubit state with phase information. The paper
is organized as follows. In Sec. II, a schematic
configuration of the Pegg-Phillips-Barnett \linebreak%
 \noindent%
QSD scheme is given and theoretical background is discussed. We
will consider an ideal scheme to study the effects of beam
splitter parameters on the fidelity of the output state and the
efficiency of truncation process. In Secs. III and IV, the
proposed experimental setup is introduced, the possible sources of
error (imperfections in detectors and single-photon source) and
their effects on the preparation of the desired state are studied.
To evaluate the feasibility of the scheme, the fidelity of the
output state and the rate of preparing it are discussed. Section V
includes a comparison of fidelities for different states. In Sec.
VI, the Wigner function and its marginals for the generated states
are analyzed. Finally, a discussion of the results is given in
Sec. VII.

\section{Quantum scissors device:~Schematics and principles}

The basic scheme of the QSD proposed by Pegg et al. is shown in
Fig. \ref{fig01}. It consists of two beam splitters and two photon
counters. The input modes of the setup are denoted as ${\hat
a}_{1}$, ${\hat a}_{2}$, and ${\hat b}_{3}$. The actions of beam
splitters can be described as unitary transformations of the
operators in the Heisenberg picture, which can be written as
\cite{Cam89}
\begin{eqnarray}
\hat{R}_{1}\hat{a}_{1}^{\dagger}\hat{R}_{1}^{\dagger}
&=t_{1}\hat{b}_{1}^{\dagger}-r_{1}^{*}\hat{b}_{2}^{\dagger} ~, \quad
\hat{R}_{1}\hat{a}_{2}^{\dagger}\hat{R}_{1}^{\dagger}
&=r_{1}\hat{b}_{1}^{\dagger}+t_{1}^{*}\hat{b}_{2}^{\dagger},
\nonumber\\%
\hat{R}_{2}\hat{b}_{3}^{\dagger}\hat{R}_{2}^{\dagger}
&=t_{2}\hat{c}_{3}^{\dagger}-r_{2}^{*}\hat{c}_{2}^{\dagger}~, \quad
\hat{R}_{2}\hat{b}_{2}^{\dagger}\hat{R}_{2}^{\dagger}
&=r_{2}\hat{c}_{3}^{\dagger}+t_{2}^{*}\hat{c}_{2}^{\dagger}, \label{N01}
\end{eqnarray}
where $\hat{R}_1$ and $\hat{R}_2$ are the unitary operators
satisfying $\hat{R}_{1}|00\rangle_{(a_{1},a_{2})}=|00\rangle
_{(b_{1},b_{2})}$ and $\hat{R}_{2}|00\rangle_{(b_{2},b_{3})}
=|00\rangle_{(c_{2},c_{3})}$. $t_{j}$ and $r_{j}$ are the
beam-splitter complex transmission and reflection coefficients
satisfying $|t_{j}|^{2}+|r_{j}|^2=1$, and $(^{*})$ denotes complex
conjugation.

In the QSD scheme, BS1 is fed by a single photon in mode ${\hat
a}_{1}$ and mode ${\hat a}_{2}$ is left in vacuum. Using the
relations given in Eq. (\ref{N01}), the output of the beam
splitter is found to be an entangled state that can be written as
\begin{eqnarray}
|\psi\rangle_{(b_{1},b_{2})} =
\hat{R}_{1}|10\rangle_{(a_{1},a_{2})}=t_{1}
|1\rangle_{b_{1}}|0\rangle_{b_{2}} - r^{*}_{1}
|0\rangle_{b_{1}}|1\rangle_{b_{2}}. \label{N02}
\end{eqnarray}
The output mode ${\hat b}_{2}$ is then fed into BS2 where it is
mixed with mode ${\hat b}_{3}$ prepared in a coherent state
\begin{eqnarray}
|\alpha\rangle_{b_{3}}=e^{-|\alpha|^{2}/2}
\sum_{n=0}^\infty\frac{\alpha^{n}}{n!}(\hat{b}_{3}^{\dagger})^{n}
|0\rangle_{b_{3}}, \label{N03}
\end{eqnarray}
which will be truncated to prepare the desired superposition of
vacuum and one-photon states
\begin{eqnarray}
|\varphi_{\rm desired}\rangle_{b_{1}}
=\frac{1}{\sqrt{1+|\alpha|^2}} (|0\rangle_{b_{1}}+
\alpha|1\rangle_{b_{1}}). \label{N04}
\end{eqnarray}
As a result of the action of BS2 with $t_{2}$ and $r_{2}$, the
state at three modes becomes
\begin{eqnarray}
&&|\Psi\rangle_{(b_{1},c_{2},c_{3})} = e^{-|\alpha|^{2}/2}
\sum_{n=0}^\infty\sum_{k=0}^n
\frac{\alpha^{n}(-r^{*}_{2})^{k}(t_{2})^{n-k}}{\sqrt{k!(n-k)!}}
\nonumber\\
&&\quad\times(t_{1}|1,k,n-k\rangle -\sqrt{(n-k+1)}r^{*}_{1}r_{2}|0,k,n-k+1\rangle
\nonumber\\
&&\quad-\sqrt{(k+1)}r^{*}_{1}t^{*}_{2}|0,k+1,n-k\rangle). \label{N05}
\end{eqnarray}
Both output modes of BS2 are measured with photon-counting
detectors. The output state generated at mode ${\hat b}_{1}$ of
BS1 depends on the outcome of the measurements at the detectors. A
normalized superposition of zero- and one-photon states
\begin{eqnarray}
|&&\Phi_{\rm trunc}\rangle_{b_{1}} = {\cal N} \,_{(c_2, c_3)}
\langle 1,0| \Psi\rangle_{(b_{1},c_{2},c_{3})}
\nonumber \\
&&\;\;\;=\frac{1}{\sqrt{|r_{1}t_{2}|^{2}+|\alpha|^{2}
|r_{2}t_{1}|^{2}}}[(r_{1}t_{2})^{*}|0\rangle_{b_{1}}+r^{*}_{2}
t_{1}\alpha|1\rangle_{b_{1}}],\label{N06}
\end{eqnarray}
where ${\cal N}$ is the renormalization constant, can be obtained
at the output of the QSD upon detection of one photon at \D2 and
no photon at \D3.

\begin{figure*}[h]
\hspace*{2.5cm}{\bf (a)}\hspace{4cm}{\bf (b)} \vspace{0mm}
\hspace*{-2mm} \epsfxsize=4.3cm
\epsfbox{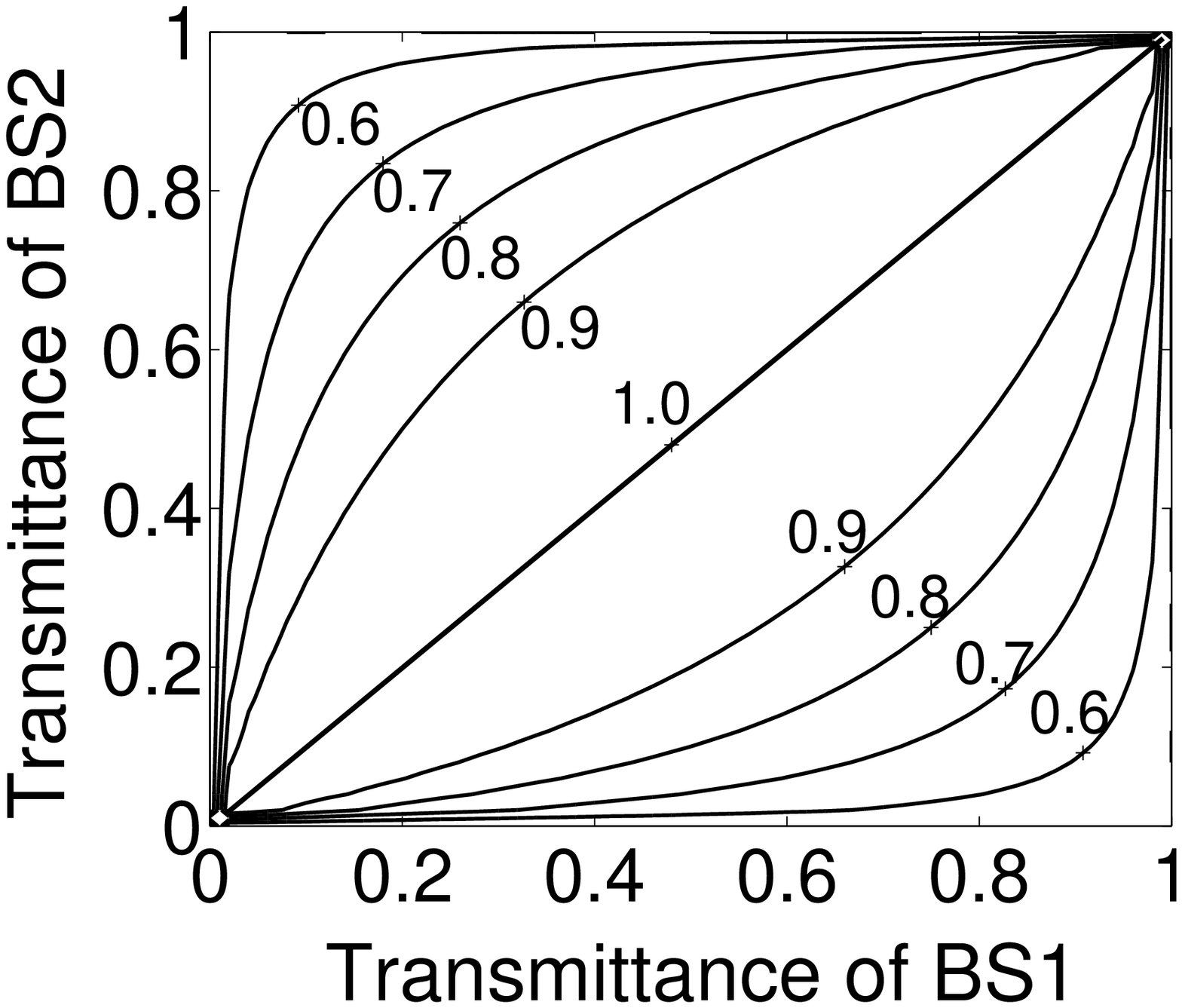}\hspace*{-0.5mm}
\epsfxsize=4.45cm \epsfbox{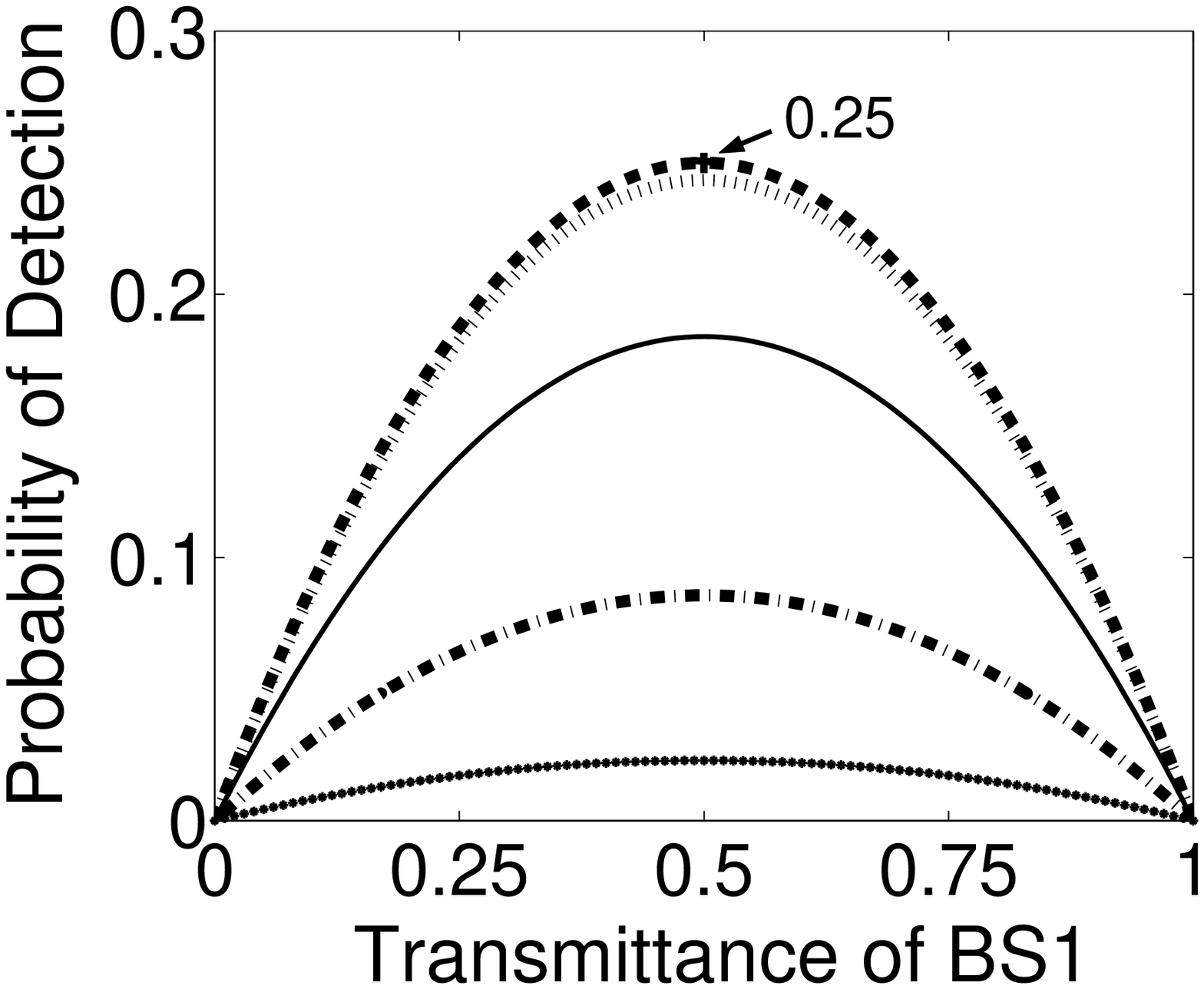} \vspace*{3mm}%
\caption{Effect of beam-splitter parameters (transmittances
$|t_{1}|^{2}$ and $|t_{2}|^{2}$) and intensity of the input
coherent light on (a) the fidelity and (b) efficiency of
truncation process (probability of proper detection). Curves (a)
are plotted for constant fidelity. In (b), beam splitters are
considered to be the same (thus $F=1$) and curves from top to
bottom correspond to $|\alpha|^2= 0.1,~0.5,~1.0,~1.5,$ and $2.0$,
respectively. The highest probability of detection is $\sim 0.25$.
}\label{fig02}
\end{figure*}

Although this scheme can also be used to obtain any desired
superposition of vacuum and single-photon states by proper choice
of beam-splitter parameters and $\alpha$, we will consider only
the truncation process in this study. The fidelity of the output
truncated state to any desired state can be calculated from
\begin{equation}\label{N07}
  F=~_{b_{1}}\langle\varphi_{\rm desired}
  |\hat{\rho}_{\rm trunc}|\varphi_{\rm desired}\rangle_{b_{1}}
\end{equation}
with $\hat{\rho}_{\rm trunc}= |\Phi_{\rm trunc}\rangle
_{b_{1}b_{1}}\langle\Phi_{\rm trunc}|$. Then the fidelity of
preparing the truncated coherent state up to single-photon state
can be found as
\begin{eqnarray}
F=\frac{|r_{1}t_{2}|^{2}+|\alpha|^{2}(r_{1}^{*}r_{2}t_{1}^{*}t_{2}^{*}
+r_{1}r_{2}^{*}t_{1}t_{2})+|\alpha|^{4}|t_{1}r_{2}|^{2}}
{(|r_{1}t_{2}|^{2}+|\alpha|^{2}|t_{1}r_{2}|^{2})(1+|\alpha|^2)},
 \label{N08}
\end{eqnarray}\noindent %
which shows that the fidelity of the truncation process depends on
the beam splitter-parameters and the intensity of the input
coherent light. Without loss of generality, we can take $r_{1}=\I
|r_{1}|$, $r_{2}=\I |r_{2}|$, $t_{1}=|t_{1}|$, and $t_{2}=|t_{2}|$
for which
\begin{equation}\label{N09}
|\Phi_{\rm trunc}\rangle_{b_{1}} =
\frac{|r_{1}t_{2}||0\rangle_{b{1}}
+\alpha|r_{2}t_{1}||1\rangle_{b{1}}}{\sqrt{|r_{1}t_{2}|^{2}
+|\alpha|^{2} |r_{2}t_{1}|^{2}}}
\end{equation}
is obtained. In that case, the dependence of truncation fidelity
on beam-splitter parameters for an input coherent light of
$|\alpha|^{2}=1$ will be as shown in Fig. \ref{fig02}(a). It can
be seen from this figure that perfect fidelity $(F=1)$  is
achieved for a range of beam-splitter parameters satisfying
$|t_{1}|^{2}-|t_{2}|^{2}=0$. However, the efficiency of
truncation, which can be defined as the probability of the desired
detection, is different for different choices of beam-splitter
parameters and can be calculated as
\begin{eqnarray}
P_{\rm detection}&=& |\,_{(c_2, c_3)}\langle 1,0|
\Psi\rangle_{(b_{1},c_{2},c_{3})}|^{2} \nonumber\\
&=&(|r_{1}t_{2}|^{2}+|\alpha|^{2} |t_{1}r_{2}|^{2})
e^{-|\alpha|^{2}}\equiv {\cal N}^{-2}.\label{N10}
\end{eqnarray} \noindent %
\noindent%
$P_{\rm detection}$ is depicted in Fig. \ref{fig02}(b) from which
it can be concluded that the highest $P_{\rm detection}$ for $F=1$
is achieved when two identical 50:50 beam splitters are used. A
fidelity of $F=1$ is achieved with $\max(P_{\rm detection})=0.184$
when $|\alpha|^{2}=1$ and $|t_{1}|^{2}=|t_{2}|^{2}=0.5$.

Further analysis of Eq. (\ref{N05}) shows that detection of one
photon at \D3 and no photon at \D2 will yield the following
truncated state:
\begin{eqnarray}
|\Phi_{\rm trunc}^{'}&&\rangle_{b_{1}} = {\cal N}\,' \,_{(c_2,
c_3)} \langle 0,1|\Psi\rangle_{(b_{1},c_{2},c_{3})}
\nonumber \\
=&&\frac{1}{\sqrt{|r_{1}r_{2}|^{2}+ |\alpha|^{2}
|t_{1}t_{2}|^{2}}}(r^{*}_{1}r_{2}|0\rangle_{b_{1}}
-t_{1}t_{2}\alpha|1\rangle_{b_{1}}),
 \label{N11}
\end{eqnarray}
where ${\cal N}\,'$ is the renormalization constant. Substituting
imaginary reflection and real transmission coefficients as above
will give a superposition state for which the relative phase
between $|0\rangle_{b_{1}}$ and $|1\rangle_{b_{1}}$ components is
$\pi$ shifted from that of Eq. (\ref{N09}). This phase shift can
be corrected by a unitary transformation of Pauli operator
$\sigma_{z}$ after the detection. Then the use of beam splitters
with parameters satisfying $|t_{1}|^{2}+|t_{2}|^{2}=1$ will give
an output state with $F=1$. For this detection case, too, the
highest probability of generating an output state with perfect
fidelity is obtained for beam splitters with
$|t_{1}|^{2}=|t_{2}|^{2}=0.5$. In fact, under these conditions,
if $\sigma_{z}$ rotation is allowed, a successful truncation is
possible when the number of photons detected at \D2 and \D3
differs by unity \cite{Bar99}.

\section{Experimental scheme for a realistic QSD}

\begin{figure*}[h]
\hspace*{-1.5cm} \epsfxsize=9.3cm
\epsfbox{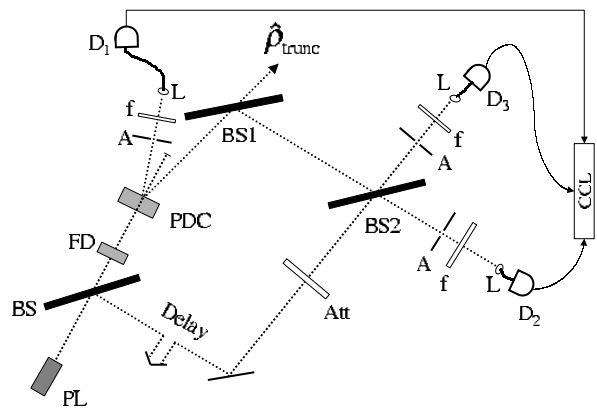}\vspace*{-17mm}%
\caption{Proposed experimental scheme for the QSD. PL, pulsed
laser; FD, frequency doubler; PDC, parametric down conversion
crystal; Att, strong attenuator; A, aperture; f, narrow band
filter; L, lens; CCL, coincidence counter and logic; BS, BS1, and
BS2 are beam splitters; and \D1, \D2, and \D3 are photon-counting
detectors.}\label{fig03}
\end{figure*}

We propose the scheme of the realistic QSD, given in Fig.
\ref{fig03}, that can be implemented in practice. One part of this
scheme uses the ideas developed and illustrated by Rarity and
co-workers \cite{Rarity2000}. The output light of a pulsed laser
with angular frequency $\omega_{0}$ is divided into two by a beam
splitter. Transmitted part of the light is frequency doubled in a
nonlinear crystal and the resultant pulses of frequency
$2\omega_{0}$ are used to pump a nonlinear crystal to induce
spontaneous parametric down-conversion (SPDC). The crystal is for
type-I degenerate phase matching, which produces down-converted
photon pairs in two modes (idler and signal) with the same
polarization and at roughly half the frequency of the pumping
pulses on opposite sides of a cone whose opening angle depends on
the angle between the optical axis of the crystal and the pump.
The pump field at the output of the crystal is eliminated by a
beam-stopping mirror. The signal and idler photons are selected by
the apertures and directed to narrow-band filters where the
background radiation is eliminated and the selection is further
restricted to only the degenerate photons. The selected idler mode
${\hat c}_{1}$ is directed to the first photon-counting detector
\D1, which is considered as a gating detector, where a ``click''
upon the detection of a photon in idler mode ensures the presence
of another photon in the signal mode ${\hat a}_1$. The latter is
input to the (50:50) BS1 at mode ${\hat a}_{1}$ and mixed with
vacuum at mode ${\hat a}_{2}$ resulting in an entangled state at
the output of BS1. The undoubled laser light beam (reflected
portion at BS) is attenuated and directed to the input mode ${\hat
b}_{3}$ of the (50:50) BS2. Then it is mixed with the entangled
state at the output mode of BS1, which is fed into the other input
mode of BS2. Temporal overlapping of these two inputs at BS2 can
be satisfied by adjusting the variable delay placed in the path of
the weak coherent state. The resultant states at the output modes
of BS2 are passed through apertures and narrow-band filters before
reaching the photon-counting detectors. Detection of a photon at
\D2 of mode ${\hat c}_{2}$ and no photons
at \D3 \linebreak%
 \noindent%
of mode ${\hat c}_{3}$ will ensure the preparation of the desired
truncated state at the output mode ${\hat b}_{1}$ of BS1. The
filters and apertures in the scheme are used to make the weak
coherent light indistinguishable from the entangled state entering
into the other input mode of BS2. In the scheme, the output state
is conditioned on coincidence detection at \D1 and \D2, and
anticoincidence at \D3.

Now, let us analyze the outlined system considering only the
effects of beam splitters and detectors. Using Eq. (\ref{N01}),
the output of a beam splitter can be calculated by
$\hat{R}\hat{\rho}_{{\rm in}}\hat{R}^{\dagger}$ for a given input
density operator $\hat{\rho}_{\rm in}$. If the output signal of
the SPDC is $\hat{\rho}_{(a_{1},c_{1})}$ then the input
state-density operator of the BS1 will be $\hat{\rho}_{{\rm
in}1}=\hat{\rho}_{(a_{1},c_{1})} \otimes |0\rangle_{a_{2} {a_{2}}}
\langle0|$. With this input, output of BS1 will be
$\hat{\rho}_{(b_{1},b_{2},c_{1})} =\hat{R}_{1}\hat{\rho}_{{\rm
in}1}\hat{R}_{1}^{\dagger} $. Considering that the state at ${\hat
b}_{3}$ mode of BS2 is coherent $|\alpha\rangle _{b_{3}b_{3}}
\langle\alpha|$, input density operator of BS2 becomes
$\hat{\rho}_{{\rm in}2}=\hat{\rho} _{(b_{1},b_{2},c_{1})}\otimes
|\alpha\rangle_{b_{3} {b_{3}}}\langle \alpha|$, then letting this
operator evolve through the BS with $\hat{R}_{2}$, we obtain the
following output density operator
\begin{eqnarray}
\hat{\rho}_{{\rm out}}&=&\hat{\rho}_{(b_{1},c_{1},c_{2},c_{3})}
= \hat{R}_{2}\hat{\rho}_{{\rm in}2}\hat{R}_{2}^{\dagger}\nonumber\\
&=& \hat{R}_{2}(\hat{\rho}_{(b_{1},b_{2},c_{1})}\otimes|\alpha
\rangle_{b_{3}}\,_{b_{3}}\!\langle\alpha|)\hat{R}_{2}^{\dagger}
\nonumber\\
&=&\hat{R}_{2}\hat{R}_{1}(\hat{\rho}_{(a_{1},c_{1})} \otimes|0
\rangle_{a_{2}}\,_{a_{2}}\!
\langle0|\otimes|\alpha\rangle_{b_{3}}\,_{b_{3}}\!\langle\alpha|)
\hat{R}_{1}^{\dagger}\hat{R}_{2}^{\dagger}.\label{N13}
\end{eqnarray}
The normalized truncated output state density operator at mode
${\hat b}_{1}$ of BS1 is obtained by
\begin{eqnarray}
\hat{\rho}_{\rm trunc}=\frac{{\rm{Tr}}_{(c_{1},c_{2},c_{3})}
(\Pi_{1}^{c_{1}} \Pi_{1}^{c_{2}} \Pi_{0}^{c_{3}} \hat{\rho}_{{\rm
out}})} {{\rm{Tr}}_{(b_{1},c_{1},c_{2},c_{3})}
(\Pi_{1}^{c_{1}}\Pi_{1}^{c_{2}} \Pi_{0}^{c_{3}} \hat{\rho}_{{\rm
out}})} \, , \label{N14}
\end{eqnarray}
where $\Pi_{1}^{c_{1}}$, $\Pi_{1}^{c_{2}}$, and $\Pi_{0}^{c_{3}}$
are elements of the positive-operator-valued measures (POVM's) for
the detectors \D1, \D2, and \D3, respectively, with zero and one
corresponding to the number of clicks recorded at the detectors.

\section{Error sources in QSD scheme}

In the following, we assume that the apertures, narrow-band
filters, and delays introduced in the scheme ensure the proper
phase and mode matching at the beam splitters. Thus, we will study
only the imperfections in the single-photon generation and
photon-counting devices, and their effects on the feasibility of
the QSD scheme.

\subsection{Nonideal single-photon source}

In the proposed scheme SPDC is used as the source for the
preparation of the single-photon state at the input of the BS1. In
practice, we must take into account some basic features of SPDC as
a source. First, although the conservation of energy forces the
sum of the frequencies of the idler and signal photons to be equal
to the frequency of the pump field, the photons may have finite
bandwidth due to the finite size of the crystal. Second is the
spatial location of the idler and signal photons in the cone of
radiation at the crystal output. These two problems can be solved
approximately by spatial and frequency filtering as explained
above. The third one is the intrinsic property of SPDC, that is,
the output of SPDC contains vacuum with high probability while the
probability of a photon-pair generation is very low. Even though
the probability is much lower, there may be cases where more than
one photon pair is generated. In SPDC, photons are generated in
pairs with equal numbers in the signal ($\hat{a}_1$) and idler
($\hat{c}_1$) modes as can be seen in the expression for the state
at the output of the SPDC crystal \cite{Yurke87},
\begin{eqnarray}
| \varphi \rangle_{(a_1,c_1)}= \sqrt{1-\gamma^{2}}
\sum_{k=0}^\infty (\gamma e^{\I \theta_{\rm p}})^k |k \rangle
_{a_1}|k \rangle _{c_1} \, , \label{N15}
\end{eqnarray}
where $\gamma = \tanh(|\kappa \tau|)$ and $\gamma^{2}$, typically
$\sim 10^{-4}$/pulse \cite{Bouw97}, corresponds to the rate of
one-photon-pair generation per pulse of the pump field; $\kappa$
represents the product of coupling constant and the complex
amplitude of the pump field; and $\tau$ stands for the interaction
time of the pump field and the crystal. The phase of the pump
field is denoted by $\theta_{\rm p}$. In the proposed experimental
scheme, we suppose that the explicit information of the phase
$\theta_{\rm p}$ is not known, which results in the following
mixed state for the output of SPDC after averaging over all
possible phases:
\begin{eqnarray}
\hat{\rho}_{(a_1,c_1)}&=&(1-\gamma^{2}) [ |00 \rangle\langle 00 |
+ \gamma^{2}|11 \rangle\langle 11 |
\nonumber\\
&&+\gamma^{4}|22 \rangle\langle 22| +\gamma^{6}|33 \rangle\langle
33| + \cdots]_{(a_1,c_1)} \label{N16}
\end{eqnarray}
In the following,  Eq. (\ref{N16}) will be used for numerical
simulations.

\subsection{Imperfect photon counting detectors}

Photodetection is the very basis of quantum-optical measurements.
Currently most commonly used photodetectors are avalanche
photodiodes that suffer from four main problems (a) nonunit
efficiency $(\eta\neq1)$ causing the failure of photon detection,
(b) nonzero dark count $\nu\neq0$ causing ``false alarms'' by
signal generation even when there is no photon, (c) failure to
discriminate between $n$ and $n+1$ photons if $n\geq1$, and (d)
``dead time" $\tau_{\rm dt}$ of the photon-counting detector and
the processing electronics during which detectors cannot respond
to the incoming photons.

After the arrival of the first photon to a detector, a time
duration of $\tau_{\rm dt}$ should pass for the detector to count
the next coming photons. If the arrival times of photons at the
detector are less than $\tau_{\rm dt}$,  then only one electronic
\begin{figure*}[h]
\hspace*{-5mm} \epsfxsize=4.6cm \epsfbox{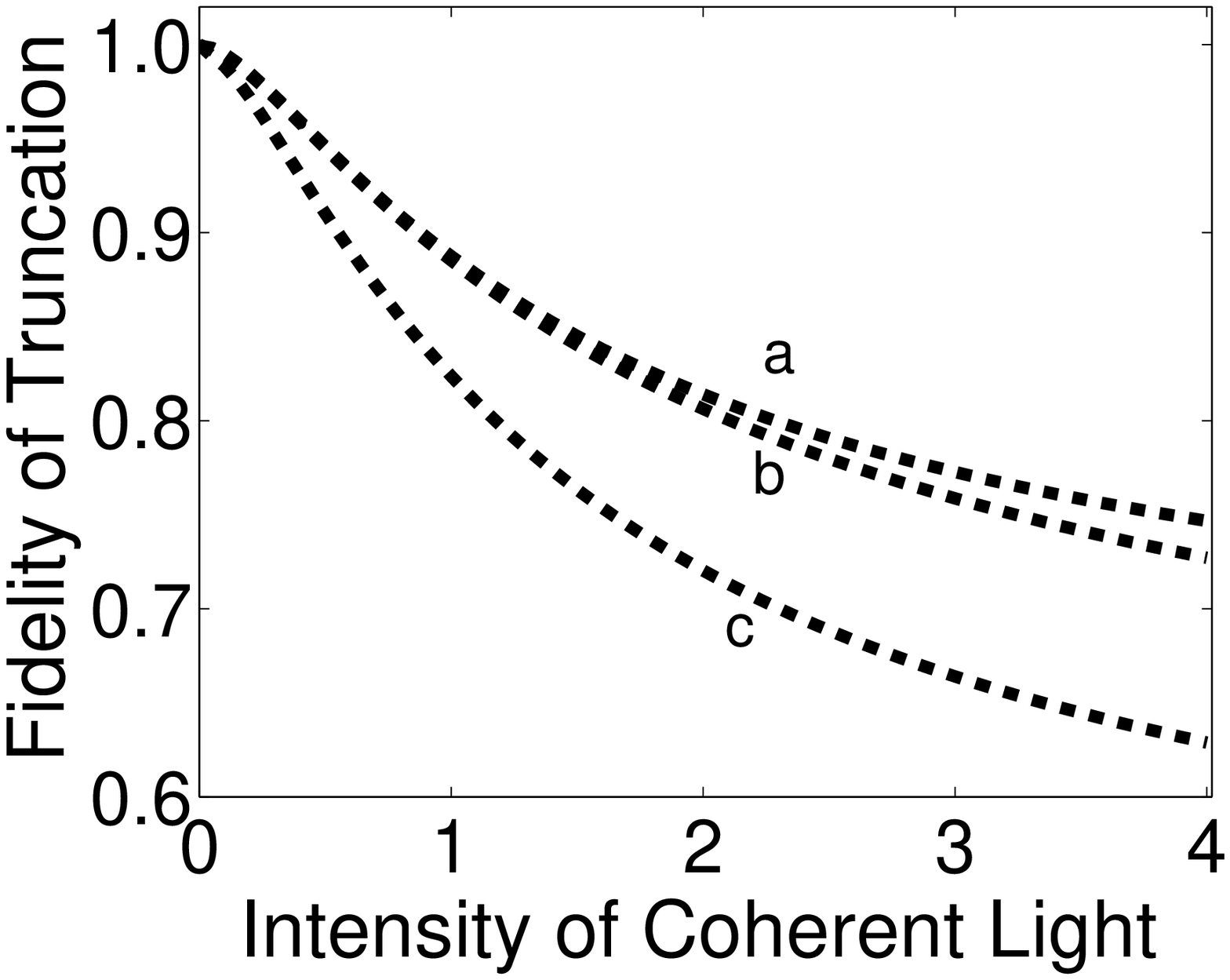}\hspace{-1mm}
\epsfxsize=3.98cm \epsfbox{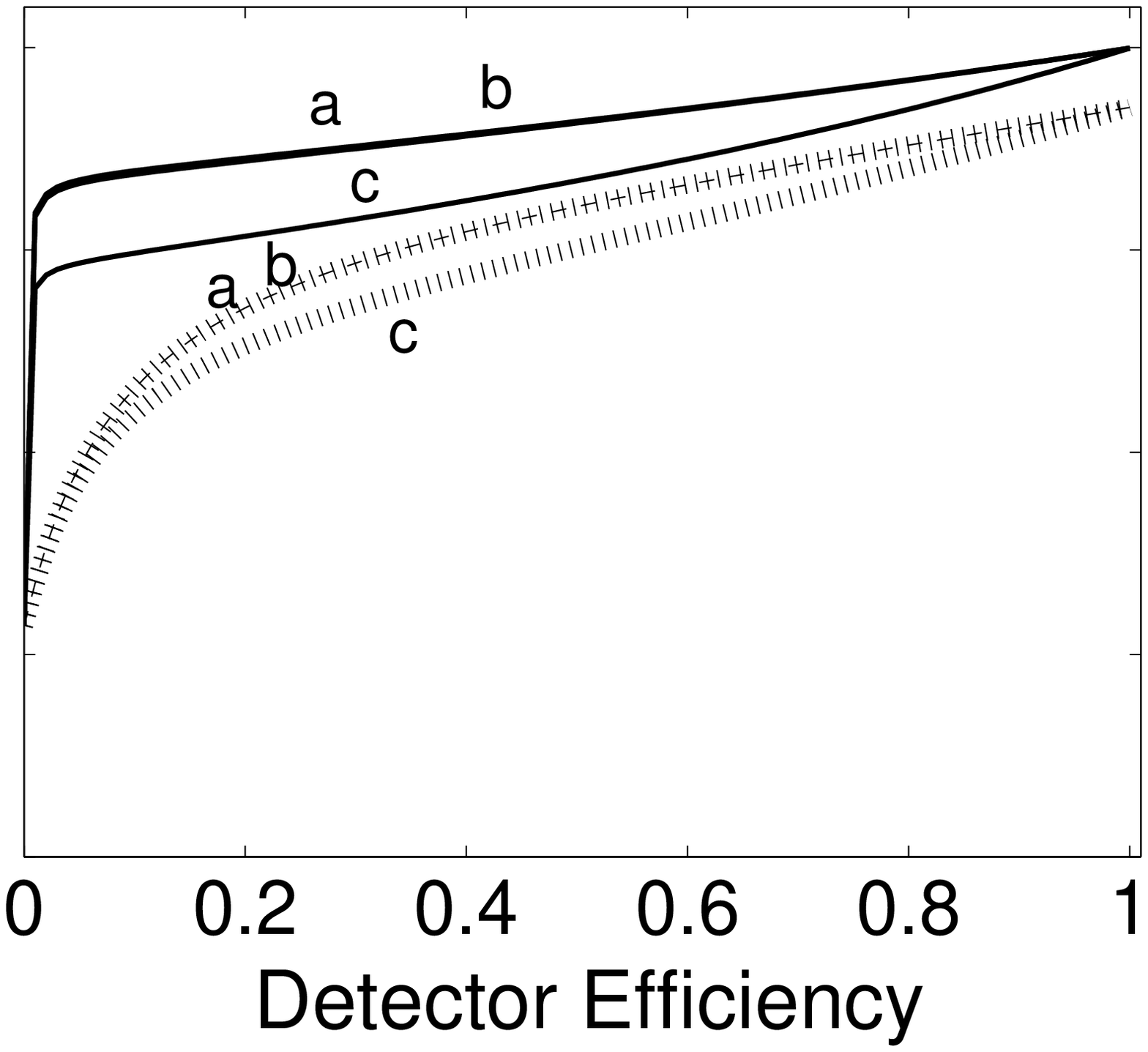}\vspace{5mm}%
\caption{Effect of intensity of coherent light, $|\alpha|^{2}$,
(left) and the detector efficiency $\eta$ (right) on the fidelity
$F$ of truncation process for photon-number-discriminating
counters and $\gamma^{2}=5\times10^{-4}$/pulse. $a$, $b$, and $c$
correspond to $(1,0)$, $(2,1)$, and $(3,2)$, respectively, which
stand for the number of photons detected at the detectors
(\D2,\D3). Left plot was obtained for $R_{\rm dark}=1000~{\rm
s}^{-1}$ and $\eta=0.5$. Right plot is for $|\alpha|^{2}=0.4$ with
solid and dotted curves corresponding to $R_{\rm dark}=100~{\rm
s}^{-1}$ and $R_{\rm dark}=10^{4}~{\rm s}^{-1}$, respectively.
}\label{fig04}
\end{figure*} \noindent %
pulse, which corresponds to the detection of the first photon,
will be generated. The average counting rate should be less than
$1/\tau_{\rm dt}$ to eliminate the effect of dead time on the
counted photon rate. In the QSD scheme, ``dead time" shows itself
in two ways. (i) For \D1 and \D2 detectors, if a photon is
incident on the detector within $\tau_{\rm dt}$ seconds after the
preceding photon, then the event will just be neglected and we
will not count the output state as the desired one. The effect of
such a case will be the reduction in the number of states
generated per second. (ii) There may be cases where \D1 and \D2
detect photons and even though there is an incident photon on \D3,
it does not click because it is still ``dead." Such a case will be
counted as the desired detection and a state different from the
desired one will be prepared at the output, which will decrease
the fidelity of the prepared states. Both of these two cases can
be considered as the loss of photon due to inefficient detection.
A decision on the outcome of the measurement is done for each
light pulse separately, independent of the result of the preceding
or the following pulses. Moreover, we are not interested in the
correlation between two consecutive pulses. Therefore, the effect
of dead time can be absorbed into the detector efficiency $\eta$.
Typical values for dead time have been reported to be in the range
30 ns $\leq\tau_{\rm dt}\leq$ 100 ns for million counts per second
\cite{CPC1}. In order to minimize the effect of dead time, one has
either to use very weak light so that the number of photons in the
system per second is low enough, or to work with small repetition
frequency for the input coherent light and the pump of the SPDC
crystal.

For a realistic description of photon-counting detectors (\D1,
\D2, and \D3) shown in Fig. \ref{fig03}, POVM can be written as
\cite{Bar98}
\begin{eqnarray}
\Pi_{N}= \sum_{n=0}^N \sum_{m=n}^\infty
\frac{e^{-\nu}\nu^{N-n}}{(N-n)!} \eta^{n}(1-\eta)^{m-n}C^{m}_{n}
|m\rangle \langle m| \label{N17}
\end{eqnarray}
for a detector with quantum efficiency $\eta$ (dead-time effects
are included) and mean dark count of $\nu$, where $\sum_{N=0}
^\infty \Pi_{N}=1$. In this equation $n$ is the actual number of
photons present in the mode, $N-n$ is the number of dark counts,
$N$ is the number of ``clicks," and $C^{m}_{n}$ is the binomial
coefficient. Mean dark count rate is given by $\nu=\tau_{\rm
res}R_{\rm dark}$, where $R_{\rm dark}$ is the dark count rate and
$\tau_{\rm res}$ is the resolution time of the detector and the
electronic circuitry and is longer than the pulse width $\tau_{p}$
of the pulsed light used in the experiment. In the following, we
will work with three kinds of detectors
(photon-number-discriminating detector, conventional photon
counter, and single-photon counter) in order to show the effects
of imperfections (a)-(c) and detector types on the properties of
truncated states using POVM for each type of detector. We will
assume a pump with a repetition frequency of 100 MHz and the
detector resolution time $\tau_{\rm res}$=10 ns.

\subsubsection{Photon-number-discriminating counter (PNDC)}

Although not available in the market, the analysis of the system
with PNDC will give us a reference for comparison with other
detector types. The POVM for this kind of detectors is given by
Eq. (\ref{N17}). For this type of detectors, if the ideal case
(unit efficiency, no dark count, and perfect single-photon source)
is considered, the same result shown in Eq. (\ref{N04}) is
obtained. In Fig. \ref{fig04}, we show the dependence of fidelity
of truncation with the experimental scheme if all three detectors
are PNDC. In this figure, we see that increasing coherent light
intensities $|\alpha|^{2}$ causes a decrease in the fidelity of
truncation for all detection cases. In the cases where $(1,0)$ and
$(2,1)$ photons are detected at detectors (\D2,\D3), fidelity of
truncation is almost the same, however, for other cases
$(n>2,~m>1)$, $F$ decreases sharply with increasing light
intensities and dark count rates. We have also observed that for
increasing light intensities, the effect of $\eta$ become very
deleterious (not shown in Fig. \ref{fig04}). However, for
$|\alpha|^{2}\ll1$, it does not constitute a major problem and the
fidelity of truncation is essentially insensitive to changes in
$\eta$.

\subsubsection{Conventional photon counter (CPC)}

\begin{figure*}[h]
\vspace*{-7mm}\hspace*{-6mm} \epsfxsize=4.3cm
\epsfbox{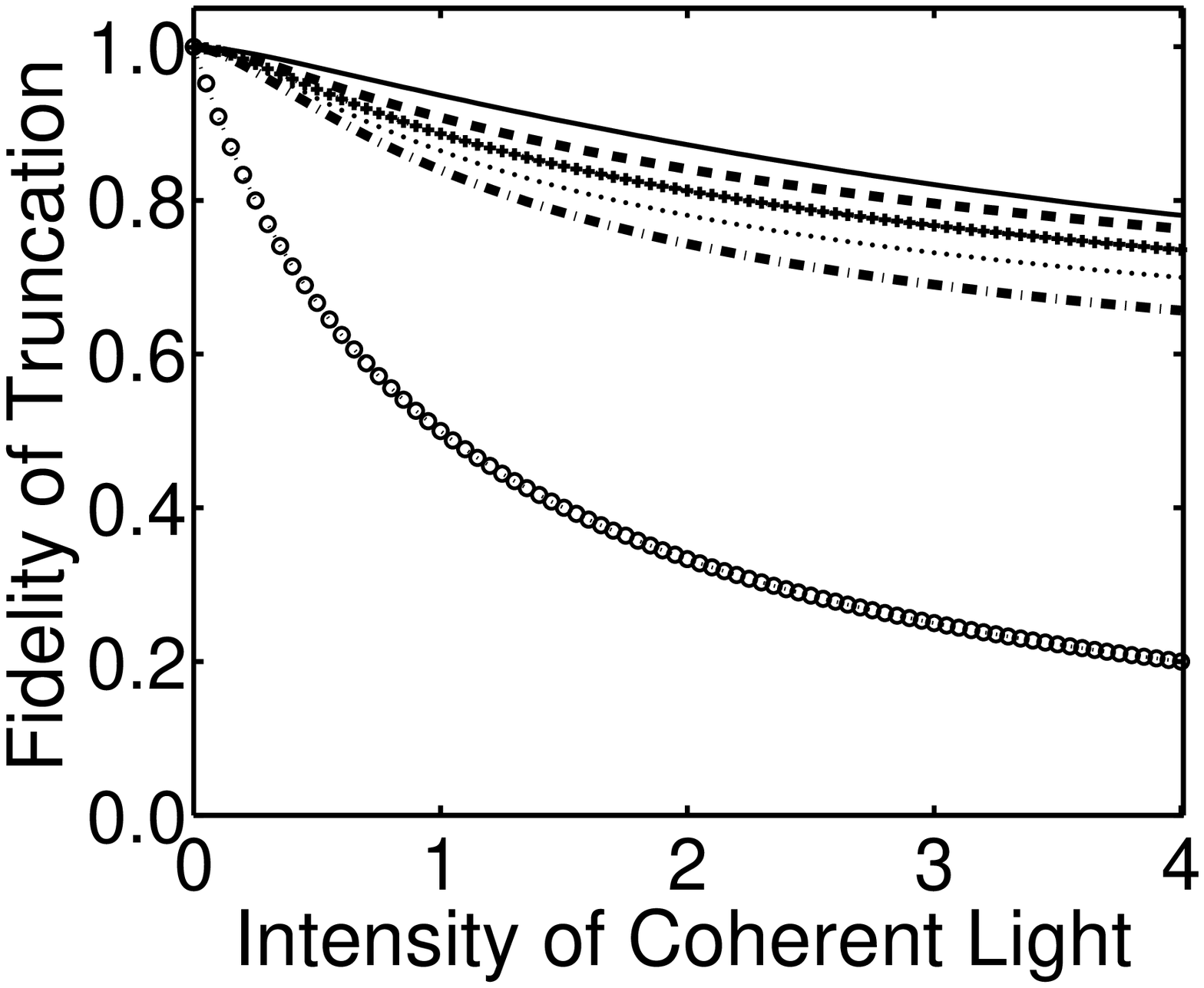}\hspace*{0mm}
\epsfxsize=4.3cm \epsfbox{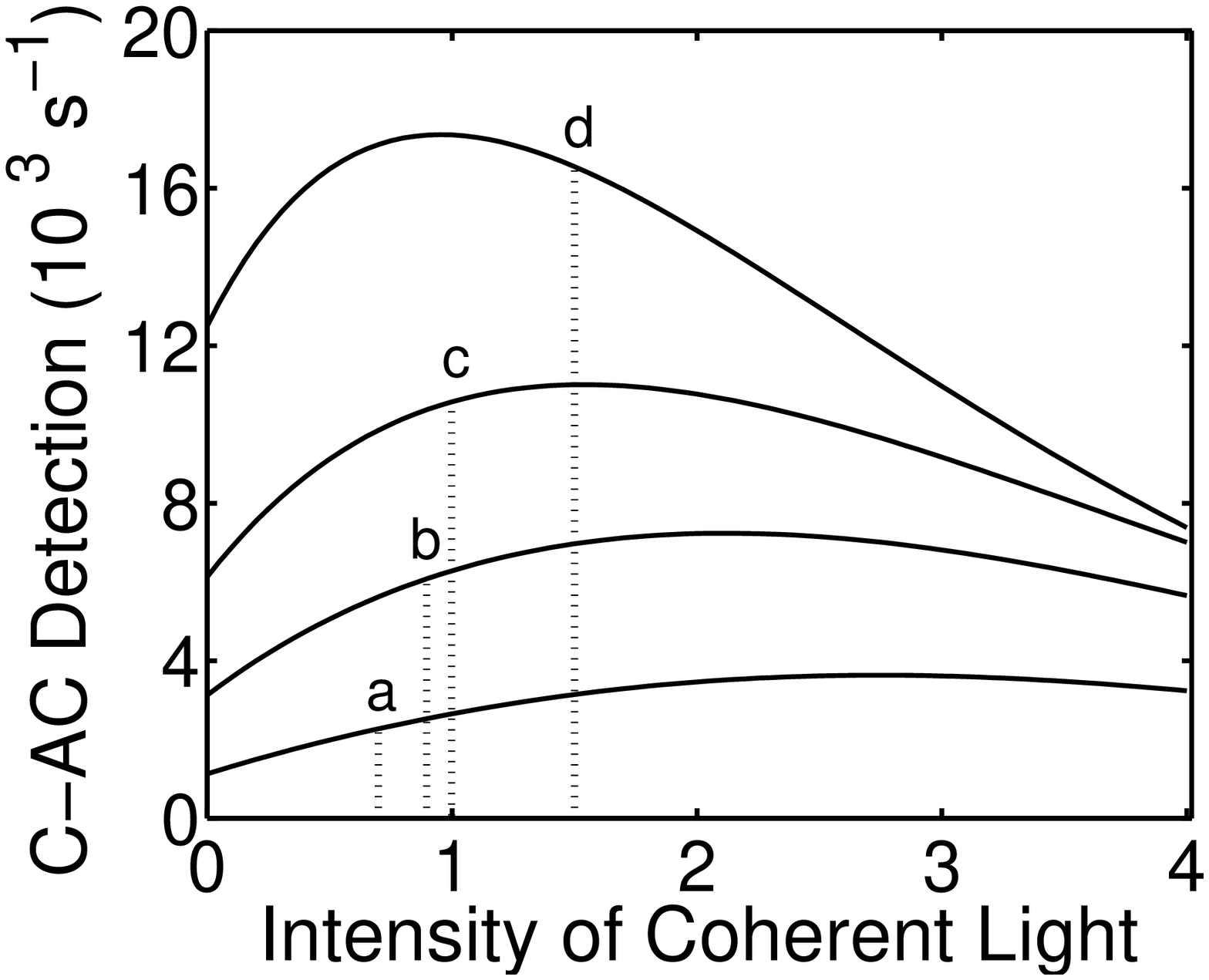} \vspace{6mm}%
\caption{Effect of intensity of the input coherent light,
$|\alpha|^{2}$, on the fidelity of truncation (left) and the
number of coincidence and anticoincidence (C-AC) detections per
second (right) when CPC's with $R_{\rm dark}=100~{\rm s}^{-1}$ are
used and $\gamma^{2}=5\times10^{-4}/$pulse. In the left plot,
$\eta=1.0,~0.7,~0.5,~0.3,~0.1$, and $0.0$ from top to bottom,
respectively. In the right plot, $\eta=0.3,~0.5,~0.7$, and $1.0$
for $a,b,c$, and $d$, respectively. }\label{fig05}
\end{figure*}

This type of detectors can only distinguish between the presence
and absence of photons in the mode by a ``click'' or ``no click."
No information on the exact number of photons can be obtained in a
single click. Then the POVM can be written as
\begin{eqnarray}
\Pi_{0} &=& \sum_{m=0}^\infty e^{-\nu}(1-\eta)^{m}|m\rangle
\langle m|,
\nonumber\\
\Pi_{N\geq1} &=& 1-\Pi_{0}\, .\label{N18}
\end{eqnarray}
These detectors are commercially available in the market with
$R_{\rm dark}\leq 100~{\rm s}^{-1}$ and $\eta\sim 0.7$
\cite{CPC1,Kwiat93}. Fig. \ref{fig05} depicts $|\alpha|^{2}$
versus fidelity and the rate of the proper detection for various
values of detector efficiency. It is seen that a fidelity $F>0.9$
is achievable for efficiencies as low as $0.7$ at
$|\alpha|^{2}=1$. Further decrease of $\eta$ from $0.7$ to $0.1$,
at this value of $\alpha$, degrades $F$ from $\sim0.92$ to
$\sim0.84$, and finally reaches $0.5$ at $\eta=0$. However, if we
restrict ourselves to work at an intensity $|\alpha|^{2}\leq0.4$,
fidelity will be higher than $0.94$ for $\eta\geq0.1$, and it will
take a value $>0.71$ for $\eta\geq0.0$. In the right plot of Fig.
\ref{fig05}, vertical dotted lines show the value of
$|\alpha|^{2}$ below which $F$ becomes $\geq0.90$ for the
corresponding detector efficiency. These values are
$0.7,~0.9,~1.0$, and $1.5$ for $\eta=0.3,~0.5,~0.7$, and $1.0$,
respectively. We have studied the effect of $R_{\rm dark}$ for
different values of $\eta$ and $|\alpha|^{2}$, and found out that
for $0~{\rm s}^2{-1}\leq R_{\rm dark}\leq1000~{\rm s}^{-1}$,
fidelity lies in the ranges $[0.99-1.0]$,~$[0.90-0.97]$, and
$[0.81-0.93]$ for $|\alpha|^{2}=0.1$, $|\alpha|^{2}=0.5$, and
$|\alpha|^{2}=1$, respectively, at $0.1\leq\eta\leq1.0$. It is
clearly understood that if the desired superposition state is to
be prepared by truncating a low-intensity coherent light, then
$R_{\rm dark}$ and $\eta$ of a commercially available CPC have
little effect on the fidelity of truncation.
\subsubsection{Single-photon counter (SPC)}

Within the context of this study, SPC is considered as the photon
counting detector that can discriminate between no photon, a
single photon and higher number of photons in the detection mode.
SPC lacks the ability to distinguish two photons from higher
number of photons. Therefore, POVM can be given as
\begin{eqnarray}
\Pi_{0}&=& \sum_{m=0}^\infty e^{-\nu}(1-\eta)^{m}|m\rangle \langle
m|,
\nonumber\\
\Pi_{1}&=& \sum_{n=0}^1 \sum_{m=n}^\infty
e^{-\nu}\nu^{1-n}\eta^{n}m^{n}(1-\eta)^{m-n}|m\rangle \langle m|,
\nonumber\\
\Pi_{N\geq2}&=& 1-\Pi_{0}-\Pi_{1}\, . \label{N19}
\end{eqnarray}

\begin{figure*}[h]
\hspace*{-6mm} \epsfxsize=4.3cm \epsfbox{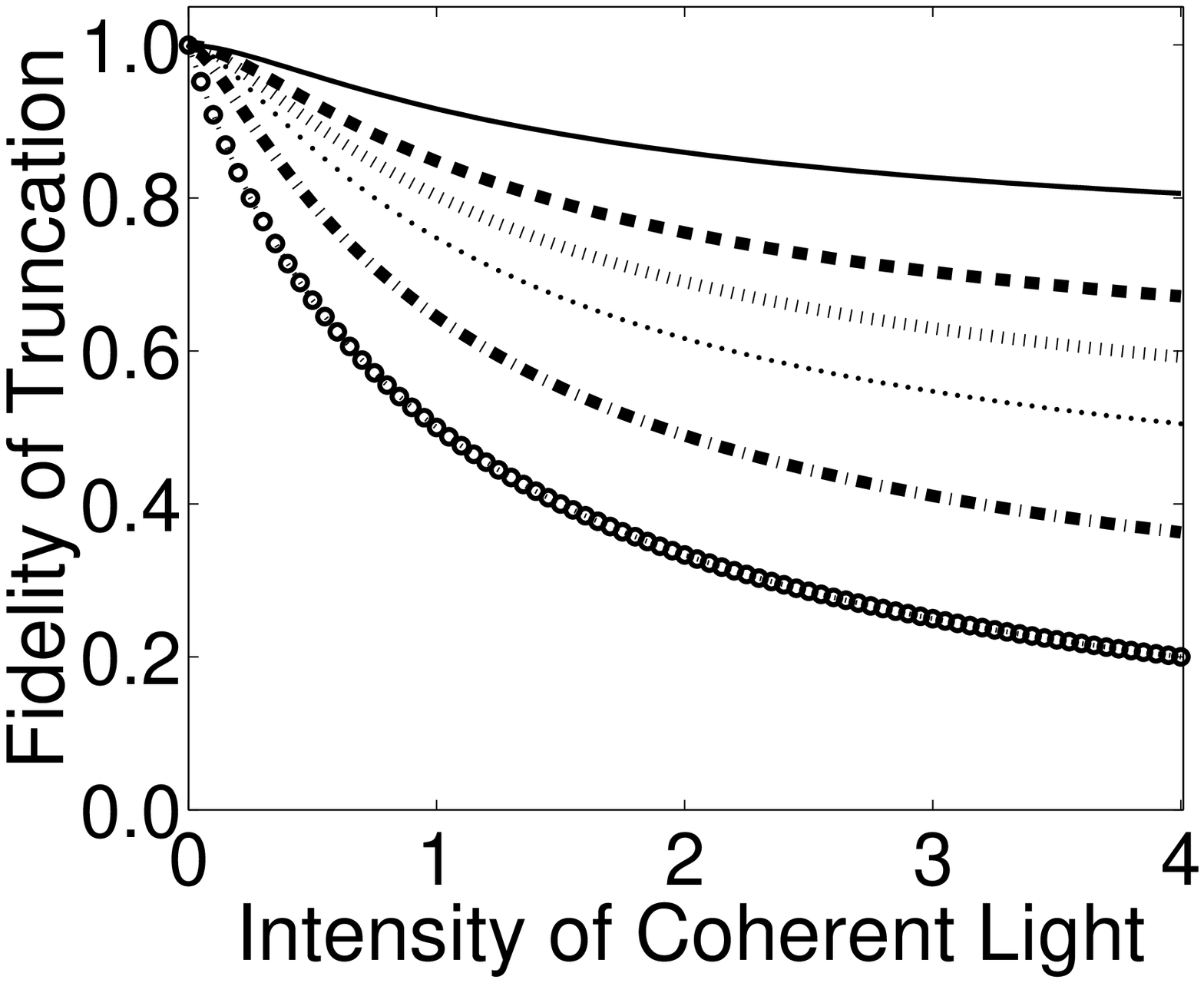}\hspace*{0mm}
\epsfxsize=4.4cm \epsfbox{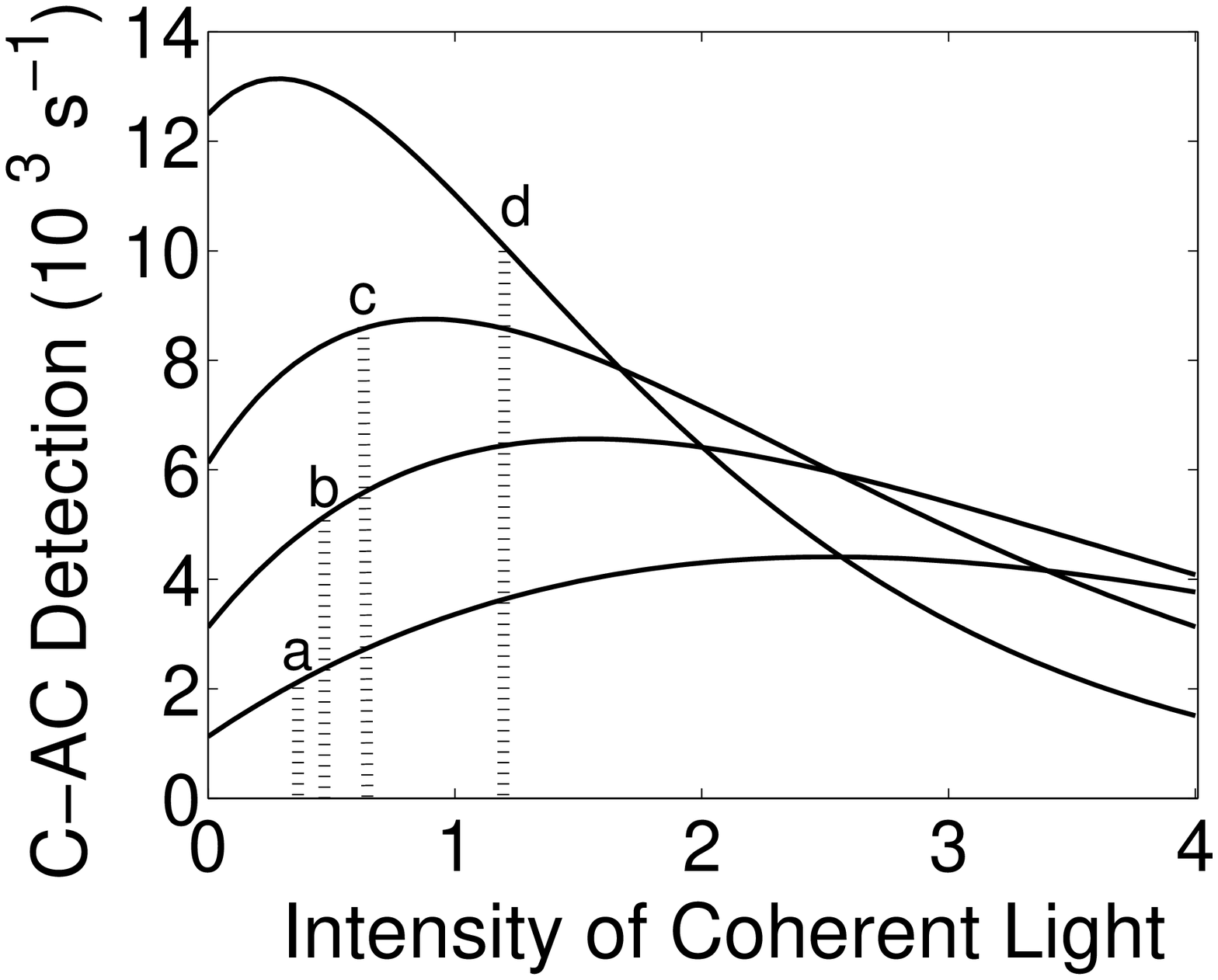} \vspace{4mm}%
\caption{Effect of intensity of the input coherent light,
$|\alpha|^{2}$, on the fidelity of truncation (left) and the
number of coincidence and anticoincidence (C-AC) detections per
second (right) when SPC's with $R_{\rm dark}=10^{4}~{\rm s}^{-1}$
are used and $\gamma^{2}=5\times10^{-4}$/pulse. In the left plot,
$\eta=1.0,~0.7,~0.5,~ 0.3,~0.1$, and $0.0$ from top to bottom,
respectively. In the right plot, $\eta=0.3,~0.5,~0.7$, and $1.0$
for  $a,~b,~c$, and $d$, respectively.}\label{fig06}
\end{figure*}

In the literature, $R_{\rm dark}\sim10^{4}~{\rm s}^{-1}$ and
$\eta\sim0.7$ have been reported for SPC's \cite{Kim99}. Figure
\ref{fig06} depicts the effect of detector efficiency and the
intensity of coherent light on the fidelity of truncation and the
number of proper detections per second when all three detectors
are SPC's. We observe that increasing coherent-light intensity
decreases fidelity; detector inefficiency is more deleterious than
the case where all detectors are CPC's. High dark count rate is a
serious problem and constitutes the main source of poor
functioning of SPC's. We have calculated fidelity of truncation
for various dark count rates at different $\eta$ and $|\alpha|^2$,
which can be summarized as follows: When $\eta=0.7$, fidelity
decreases from $0.93$ to $0.84$ if $R_{\rm dark}$ increases from
$100~{\rm s}^{-1}$ to $10^{4}~{\rm s}^{-1}$ for $|\alpha|^2=1.0$
and from $0.98$ to $0.94$ for $|\alpha|^2=0.4$. We also observed
that effect of $\nu$ is more deleterious when $\eta$ is low and
$|\alpha|^2$ is high, i.e., at $|\alpha|^2=0.4$, $F$ decreases
from $0.95$ to $0.89$ with an increase in $R_{\rm dark}$ from
$100~{\rm s}^{-1}$ to $10^{4}~{\rm s}^{-1}$, however, at
$|\alpha|^2=1.0$, it decreases from $0.87$ to $0.75$.

The desired state can be obtained any time when the difference in
the number of photons detected at \D2 and \D3 is one.
Consequently, the case when a single ``click" is detected at \D3
and two ``clicks" are detected at \D2 must also be considered. In
this case, we have seen that for the reported parameters in the
literature for SPC's, $680$ output states with $F=0.84$, and $90$
output states with $F=0.92$ at $|\alpha|^2=1.0$ and
$|\alpha|^2=0.4$, respectively, can be obtained per second. For
lower intensities the \linebreak
\vspace{5mm}
\vspace{2mm}
\noindent%
number will drop to less than four states per second, and for
higher intensities it will increase to up to $4000~{\rm s}^2{-1}$
but with much lower fidelities of around $0.6$.

\subsubsection{Discussion of different detection strategies}

In the analysis of CPC and SPC above, we have observed that CPC
has the advantage of low dark count over SPC, however, SPC has the
ability to discriminate between zero, one and more photons from
each other. The best results would have been obtained, if we had
photon counters combining these two advantages. Unfortunately, the
current level of technology does not provide this to
experimenters. However, in our QSD scheme, we can use a
combination of CPC and SPC to benefit from their unique
advantages. Simulations, with $R_{\rm dark}=100~{\rm s}^{-1}$ and
$R_{\rm dark}=10^{4}~{\rm s}^{-1}$ for CPC and SPC, respectively,
and $\eta=0.7$ for both, have shown that the choice of either CPC
or SPC for \D3 does not cause a change at the fidelity of the
output state. This can be easily understood by examining the
$\Pi_{0}$ of the detectors. Dark count for \D3 shows itself as
$e^{-\nu}$, which takes the value of $\cong1$ for both CPC and
SPC, resulting in the same value  for fidelity. Therefore, we have
only four different strategies for detector choice as shown in
Table I.

\begin{figure*}[h]
\hspace*{-4mm} \epsfxsize=4.2cm \epsfbox{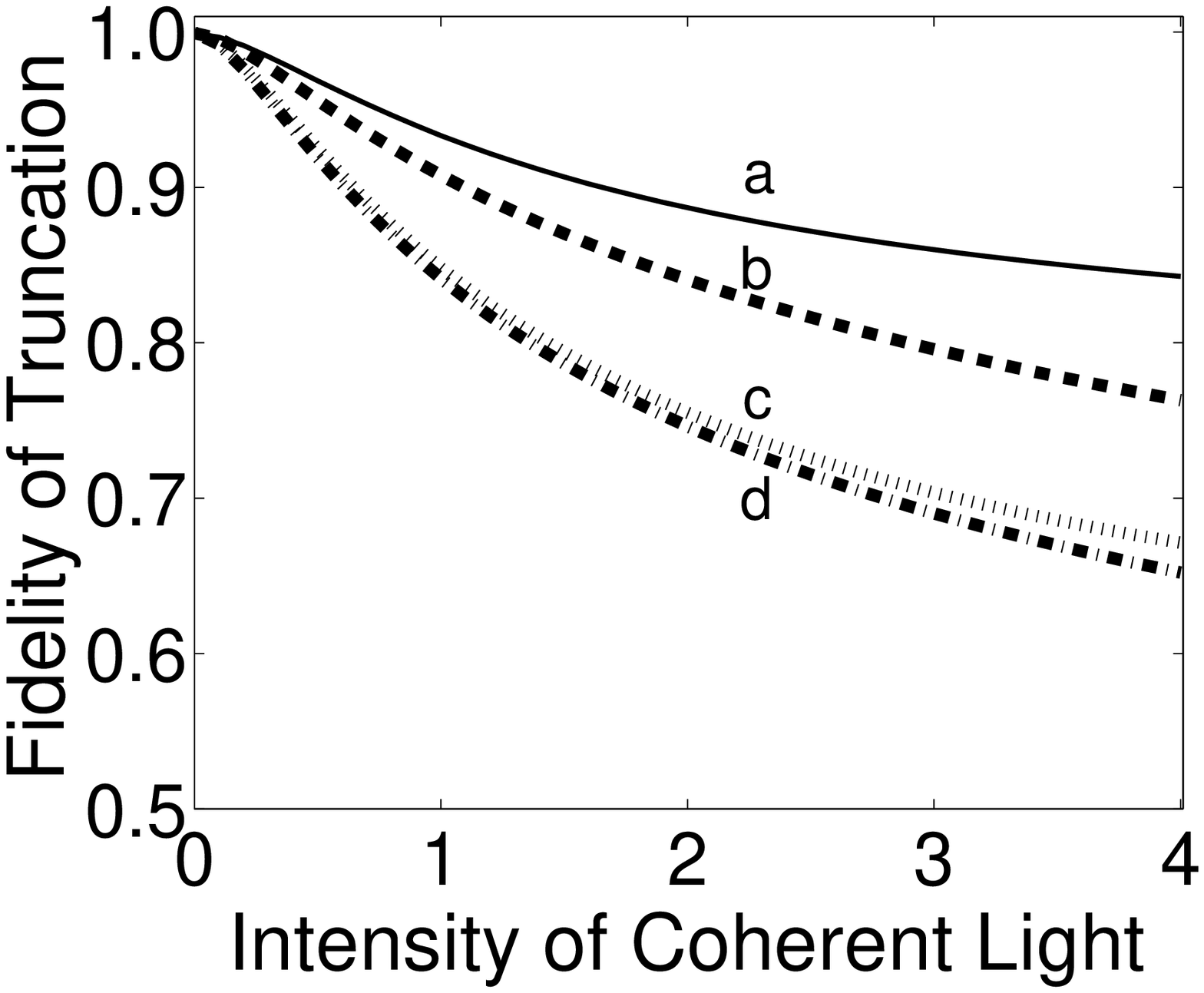}
\hspace*{0mm}\epsfxsize=4.2cm \epsfbox{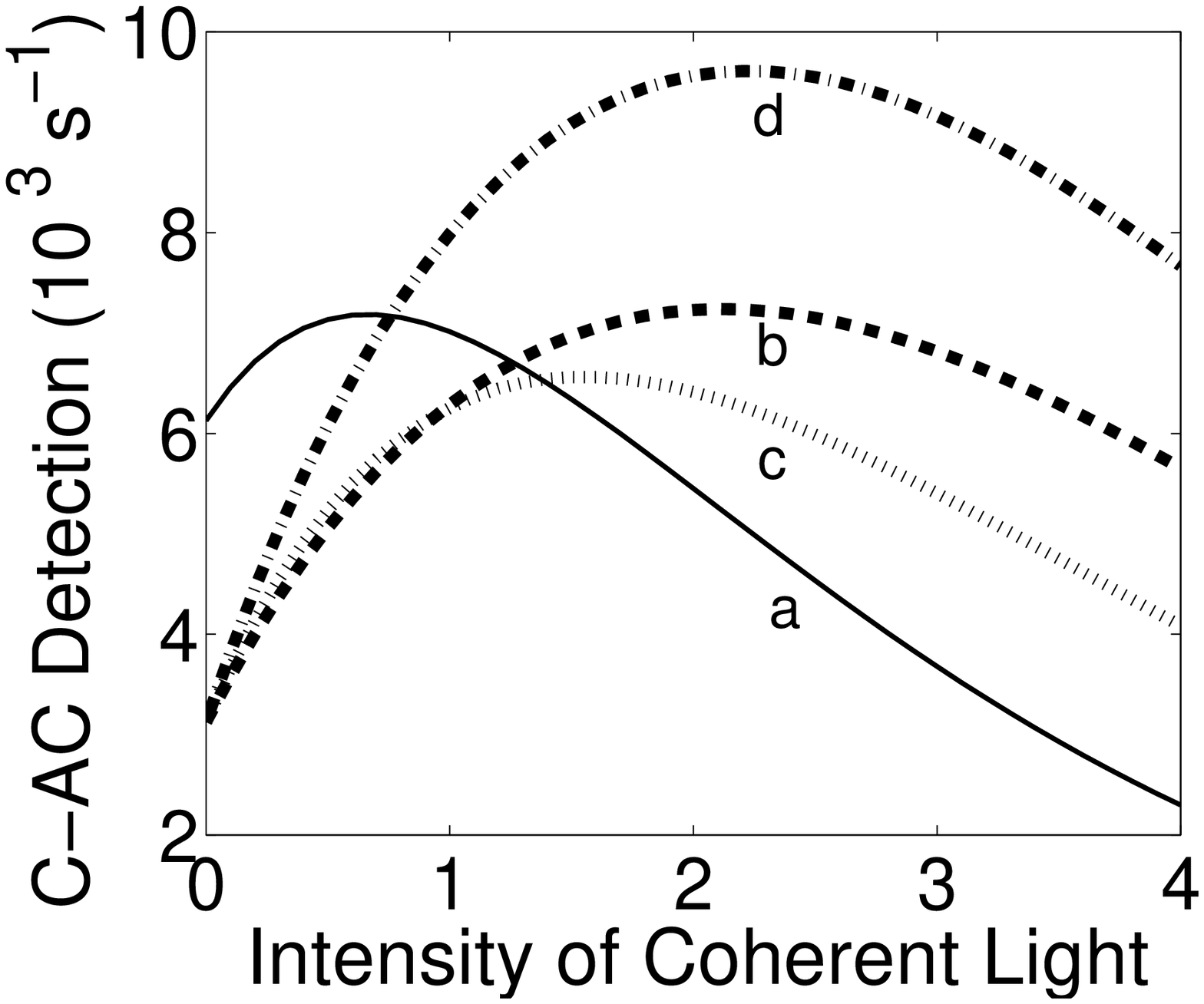} \vspace{4mm}%
\caption{Effect of intensity of the  input coherent light on the
fidelity of truncation (left) and the number of coincidence and
anticoincidence (C-AC) detections per second (right) for different
strategies of detector choice. }\label{fig07}
\end{figure*}

In Fig. \ref{fig07}, it is seen that, for
$0\leq|\alpha|^{2}\leq4$, higher fidelity values are obtained for
strategies $a$ and $b$, which use CPC's as the ``gating detector"
\D1 and much lower ones are obtained for $c$ and $d$, which use
SPC's as \D1. We explain this as follows: Probability $P_{\rm
real}$ of detecting a ``click" at \D1 caused by real photons
coming from SPDC is $O(\gamma^{2})$ per pulse, and the probability
of a ``click" caused by a dark count is $P_{\rm dark}=O(\nu)$.
Then the condition that a ``click" is caused by a real photon
rather than a dark count can be written as $P_{\rm real}\gg P_{\rm
dark}$, which implies $\nu\ll\gamma^{2}$. In the simulations, we
used a coincidence window of 10~ns and $\gamma^{2}$ of order
$\simeq10^{-4}$/pulse. Then CPC has $\nu=10^{-6}$, which satisfies
the above condition. SPC has $\nu=10^{-4}\approx\gamma^{2}$, which
means that if an SPC is used for \D1 with coincidence window of
10~ns, there will be wrong triggerings and these will reflect
themselves as decrease in fidelity. The parameter $\gamma^2$ has a
profound effect on the fidelity of truncation for the four
strategies. \linebreak
\begin{table}
\begin{tabular}{c c c c }
Strategy & \D1 & \D2 & \D3
\\\hline
a&CPC&SPC&SPC or CPC\\
b&CPC&CPC&SPC or CPC\\
c&SPC&SPC&SPC or CPC\\
d&SPC&CPC&SPC or CPC\\
\end{tabular}
\vspace{2mm}\caption{Different strategies of detector choice for a
realizable QSD. CPC, conventional photon counter; SPC,
single-photon counter; \D1, gating detector; \D2 and \D3,
detectors counting one and zero photons,
respectively.}\label{tab:table1}
\end{table}
\vspace{0mm} \noindent%
Increasing the probability of generating single-photon pair from
SPDC will increase the fidelity of truncation for those strategies
where SPC is used as \D1 and decrease that of CPC. This is because
the increasing values of $\gamma^{2}$ beyond $\nu\simeq10^{-4}$
for SPC will diminish the dark count effects.

Moreover, SPC can distinguish the number of incident photons, which will lower the probability of a ``false
alarm" due to the generation of two or more photon pairs. On the other  hand, in case of CPC, there will be
``false alarms"  that will reduce fidelity.

For $|\alpha|^{2}\leq0.6$, fidelity $F$ is $>0.90$ for all
strategies with a proper detection rate of ${\cal O} (10^{3} ~{\rm
s}^{-1})$. Then for truncating a low-intensity coherent state to
prepare the desired superposition of vacuum and single-photon
states, one can use CPC, because with a low intensity it is
guaranteed that the number of photons in the system is less than
two photons during a single preparation phase, and this will
decrease the probability of having ``false alarms" from \D2 and
\D3. By contrast, if a strong light is used then we will have
``clicks" at \D2 and \D3 when single or higher number of photons
are incident on them, and since we cannot get information on the
number of photons, we will still consider them as a sign of the
desired truncation  process, which will most of the time not be
true.

\section{Comparison of fidelities for different states}

An ideal perfect QSD scheme allows the truncation of a coherent
state up to its single-photon state with $F=1$ conserving the
relative phase and amplitude information. In an experimental
realization, $F=1$ cannot be achieved due to error sources
discussed in Sec. IV and fidelity depends strongly on the
intensity of the input coherent light. One can always ask whether
the fidelity values close to those obtained for the states
generated with the experimental scheme can be obtained for some
other states and how the state
\linebreak%
\begin{figure*}[h]
\vspace*{-2mm}\epsfxsize=5.2cm
\centerline{\epsfbox{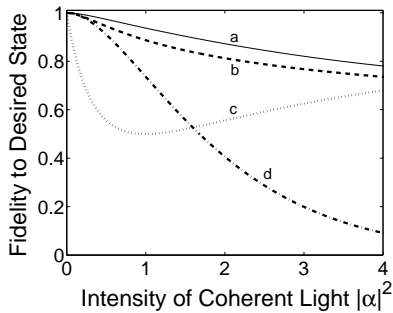}} \vspace{3mm}%
\caption{Comparison of fidelity of various states to the desired
state $\sqrt{N}(|0\rangle+\alpha|1\rangle)$ with the normalization
constant $N$. a, State obtained from experimental scheme with
$\eta=1$; b, $\eta=0.5$ (dotted curve); c,
$N(|0\rangle\langle0|+|\alpha|^{2} |1\rangle\langle1|)$; and d,
coherent state $|\beta\rangle$ with $\alpha=\beta$.
}\label{fig08} \end{figure*} \noindent %
prepared by QSD differs from those states. We will study two
cases: (i) complete loss of phase information during the
truncation process, which will yield the state $N(|0\rangle
\langle0| +|\alpha|^{2}|1\rangle\langle1|)$, where $N$ is the
normalization constant $1/(1+|\alpha|^{2})$ and (ii) a coherent
state obtained by attenuating the input $|\alpha\rangle$,
\begin{equation}\label{N20}
  |\beta\rangle=\big{|}\sqrt{\xi}~\alpha\big{\rangle} ,
\end{equation}
which will be sent directly to the output without going through
the truncation process of the QSD. Fidelities of these states to
the desired state $\sqrt{N}(|0\rangle+\alpha|1\rangle)$ are found
as
\begin{eqnarray}\label{N21}
F_{1}&=&\frac{1+|\alpha|^{4}}{[1+|\alpha|^{2}]^{2}}\, ,
\nonumber\\
F_{2}&=&\frac{e^{-|\beta|^{2}}}{1+|\alpha|^{2}}(1+2|
\alpha\beta|\cos\Delta+|\alpha\beta|^{2}),
\end{eqnarray}
where $\Delta$ is the difference of arguments of the input
coherent light $\alpha$ to be truncated and the coherent light of
$\beta$. The optimum value for $\beta$ to obtain the maximum
fidelity to the desired state for any $\alpha$ that is input to
the QSD can be found as
\begin{eqnarray}\label{N22}
|\beta|^{2}&=&\xi|\alpha|^{2}=\frac{1+2|\alpha|^{2}
-\sqrt{1+4|\alpha|^{2}}}{2|\alpha|^{2}}\, ,
  \nonumber\\
  \nonumber\\
  \arg(\beta)&=& \arg(\alpha).
\end{eqnarray}\noindent %
In Figs. \ref{fig08} and \ref{fig09}, we have depicted the
fidelities for these cases together with those of the states
obtained from the proposed experimental scheme for
$0\leq|\alpha|^{2}\leq4$. With a perfect QSD scheme and correct
information of one ``click" at \D2 and no ``click" at \D3,
fidelity is one for any $|\alpha|^2$. For the state given in case
(i), $F_{1}$ decreases gradually from $1.0$ to $0.5$ for
$0\leq|\alpha|^2\leq1.0$ and then starts increasing from $0.5$ at
$|\alpha|^{2}=1.0$ to $0.68$ at $|\alpha|^{2}=4.0$. For the
proposed scheme with CPC's as the photon-counting detectors,
fidelity
\linebreak%
\begin{figure*}[h]
\vspace*{-3mm}\hspace*{11mm} \epsfxsize=5.5cm
\epsfbox{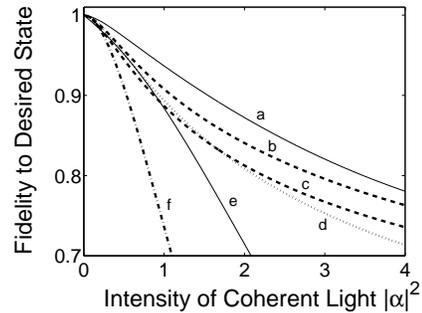} \vspace{4mm} %
\caption{Comparison of the fidelity of the states obtained from
the proposed experimental scheme of CPC's with a, $\eta=1.0$; b,
$\eta=0.7$; and c, $\eta=0.5$ with the fidelity of $|\beta\rangle$
of d, the optimized $\beta$ given by Eq. (\ref{N22}); e,
$|\beta\rangle$ with $\xi=1/2$; and f, $|\beta\rangle$ with
$\xi=1$.}\label{fig09}
\end{figure*}\noindent %
of truncation is always higher than the fidelity values of case
(i) in this range of $|\alpha|^2$ provided that $\eta\geq0.5$. For
the state given in case (ii), the fidelity for the optimized value
of $\beta$ given by Eq. (\ref{N22}) is shown in Fig. \ref{fig09}
as curve d. The states prepared with the experimental QSD scheme
have higher fidelity than the optimized $|\beta\rangle$ for
$|\alpha|^{2}\geq0.4$ at $\eta\geq0.7$. This can be observed even
for much lower $\eta$ at higher values of $|\alpha|^{2}$, i.e.,
for $|\alpha|^{2}>1.5$, $\eta=0.5$ is enough for the experimental
scheme to have better fidelity. For some values of $\xi$ such as
$\xi=1/2$ and $\xi=1$, the $|\beta\rangle$ states may have better
fidelity for low $|\alpha|^{2}$. However, with increasing
$|\alpha|^{2}$, there is a sharp monotonic decrease in their
fidelities, which finally become very close to zero ($<0.09$) for
$|\alpha|^{2}>4$. As a result, we can say that the proposed
experimental scheme is more advantageous than the other
strategies, because the states generated by the experimental
scheme have higher fidelity for a broad range of $|\alpha|^{2}$
and $\eta$. In a limited range of low $|\alpha|^{2}$, other
strategies may be optimized to give better fidelity, but only with
the cost of much lower fidelity outside this range. As we will
discuss in the next section, an analysis of the quasidistributions
of those states will give us further information to discriminate
those states from each other and to evaluate the efficiency and
the merit of using the QSD scheme.

\section{Quasidistributions for the Truncated Output State}

In the previous section, to evaluate the QSD scheme, we have used
{\it fidelity} to quantify how close the generated output state
and the desired state are. However, fidelity is just a single
number and does not give complete information on how well the
phase and amplitude of the input to the QSD are preserved at the
output. To answer this question we use the Wigner function as a
tool since it is a one-to-one representation of the quantum state
and contains all the information on state. In the following, the
Wigner function is calculated as (see, e.g., \cite{Tan96})
\begin{eqnarray}
W(X,P)=\frac{1}{\pi}\sum_{m,n}\rho_{mn}~\langle{n}| \hat{T}(X,P)
|m\rangle, \label{N23}
\end{eqnarray}
where
\begin{eqnarray}
\langle{n}|\hat{T}(X,P)|m\rangle&=&(-1)^{n}~2^{m-n+1} \, (X-\I
P)^{m-n}\sqrt{\frac{n!}{m!}}
\nonumber\\&&\times\exp(-2r^{2})~L_{n}^{m-n}(4r^{2}) \label{N24}
\end{eqnarray}
with $r^2=X^2+P^2$, $L_{n}^{m-n}(y)$ being the associated Laguerre polynomial.

\begin{figure*}[h]
\vspace*{-4mm}\hspace*{-6mm} \epsfxsize=4.5cm
\epsfbox{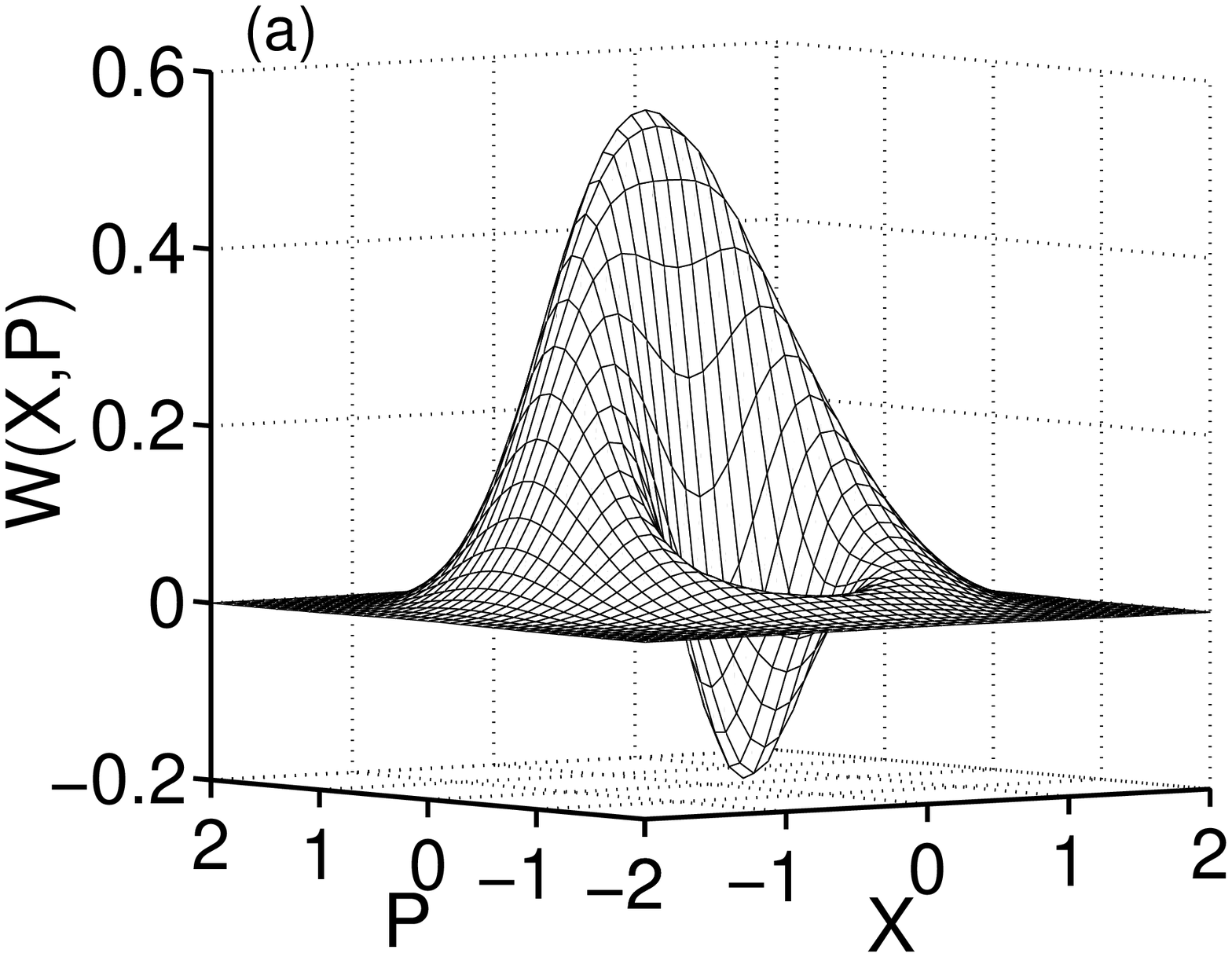}\hspace*{0mm}\epsfxsize=4.4cm
\epsfbox{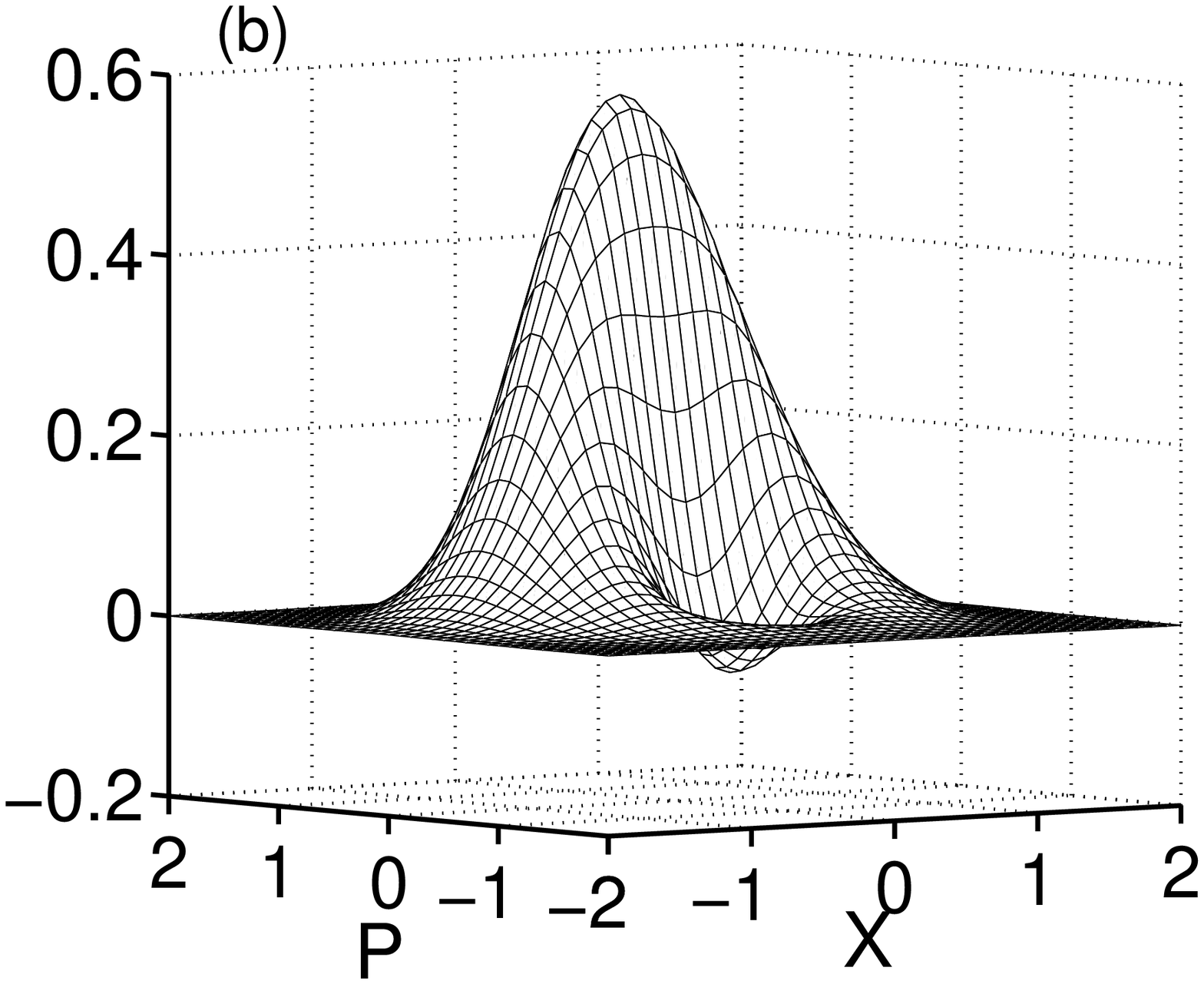} \vspace{5mm} \caption{Wigner function for
superpositions of vacuum and single-photon states obtained by
truncating coherent state $|\alpha\rangle$ with $|\alpha|^2=0.8$
and phase $\pi/2$ using (a) the perfect QSD and (b) proposed
experimental scheme with CPC's for $\eta=0.7$, $R_{\rm
dark}=100~{\rm s}^{-1}$, and $\gamma^{2}=5\times10^{-4}$/pulse.
}\label{fig10}
\end{figure*}

One of the most interesting characteristics of the QSD scheme is
the generation of nonclassical states from a classical state.
Negative values in the Wigner function are a sign of the
nonclassical property of a state. Figure \ref{fig10} shows the
Wigner functions of the desired superposition state and the
generated output state using the proposed experimental scheme with
CPC's as the photon-counting detectors. It is understood that even
for $\eta=0.5$, the aim of generating a nonclassical state is
achieved, however, we can argue that the information content
(amplitude and relative phase) of the input state is partially
lost during the truncation process due to the losses in the
system. With decreasing $\eta$, the negativity in $W(X,P)$ becomes
smaller and it is completely lost with further decrease beyond
$\eta<0.4$.

If one considers the use of the optimized $|\beta\rangle$ state
given by Eq. (\ref{N22}) rather than the QSD scheme to generate a
state with the highest fidelity to the desired state of the form
$\sqrt{N}(|0\rangle+\alpha|1\rangle)$ with $|\alpha|^{2}=0.8$ as
in Fig. \ref{fig10}, an attenuation of $\xi\sim 0.43$, which will
give an intensity of $|\beta|^{2}\sim 0.34$, must be used. In that
case, fidelity to the desired state will be $0.92$, which is
higher than the fidelity value of $F\sim 0.91$ obtained at
$\eta=0.5$ and slightly lower than $F\sim 0.93$ obtained at
$\eta=0.7$ if the proposed experimental scheme with CPC's is used.
Looking at only the values of fidelity of those states, one can
conclude that it is difficult to discriminate these states from
each other and may underestimate the advantage of using the QSD
scheme. However, if the Wigner functions of those states are
compared, the difference will become clearer. The Wigner function
for $|\beta\rangle$ with $|\beta|^{2}\sim 0.34$ will be similar to
Gaussian $W(X,P)$ for a coherent state with the peak located at
$\sim 0.34$7 and having a circular symmetry where one cannot
observe the negativity and the deformation seen in the desired
state given in Fig. \ref{fig10}(a). On the other hand, although
the fidelity values are very close to those of $|\beta\rangle$,
the states generated by the QSD scheme have deformation and
negativity similar to that of the desired state as seen in Fig.
\ref{fig10}(b).

There is a delicate balance between the intensity of the coherent
light and the efficiency of detectors to observe the negativity in
$W(X,P)$. To show this clearly, we have analyzed marginal
distributions and cross sections
\linebreak%
\begin{figure*}
\vspace{0mm} \hspace*{0mm} \epsfxsize=4cm
\epsfbox{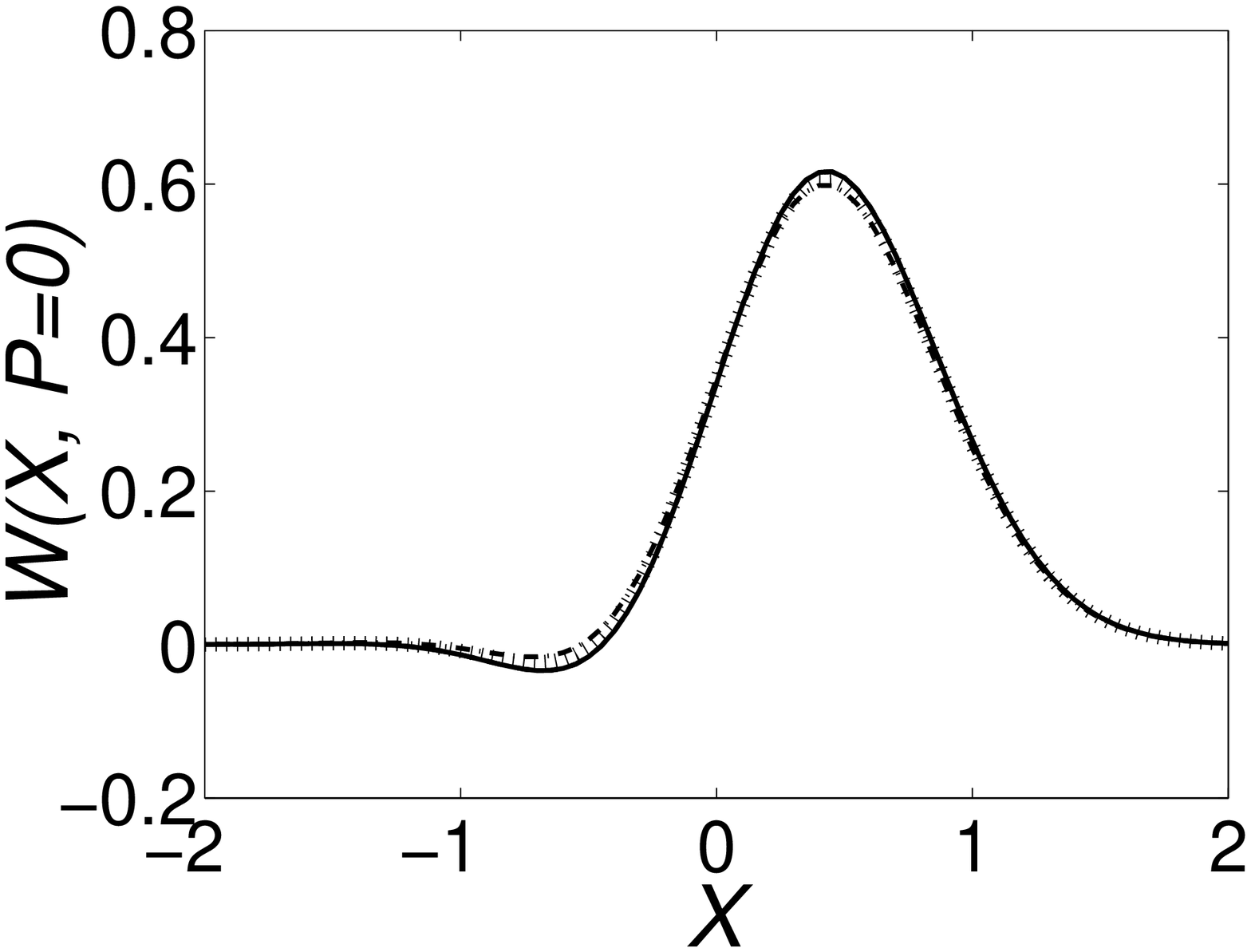}\hspace*{0mm}\epsfxsize=3.6cm
\epsfbox{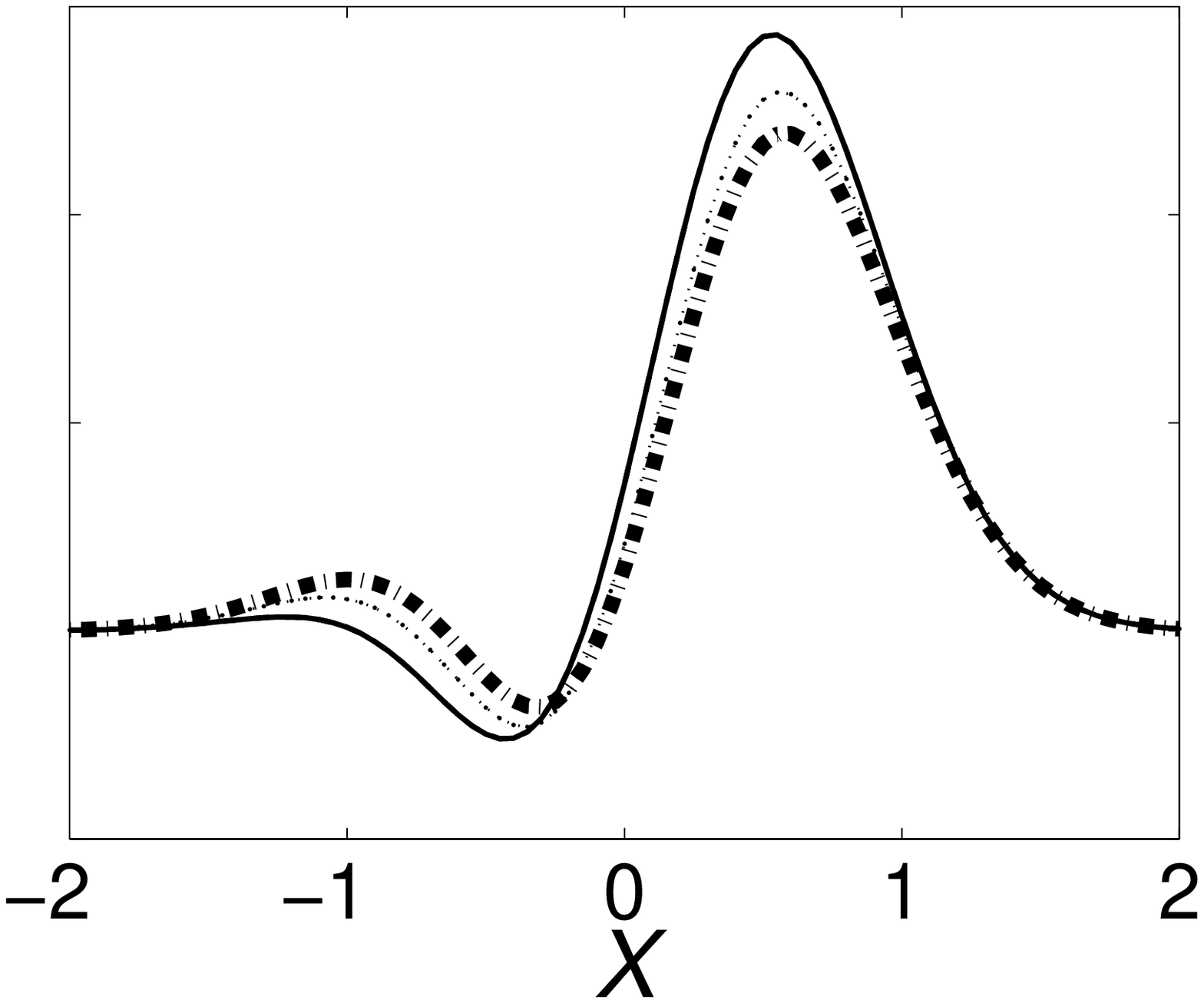} \\
\hspace*{5mm}\epsfxsize=4cm
\epsfbox{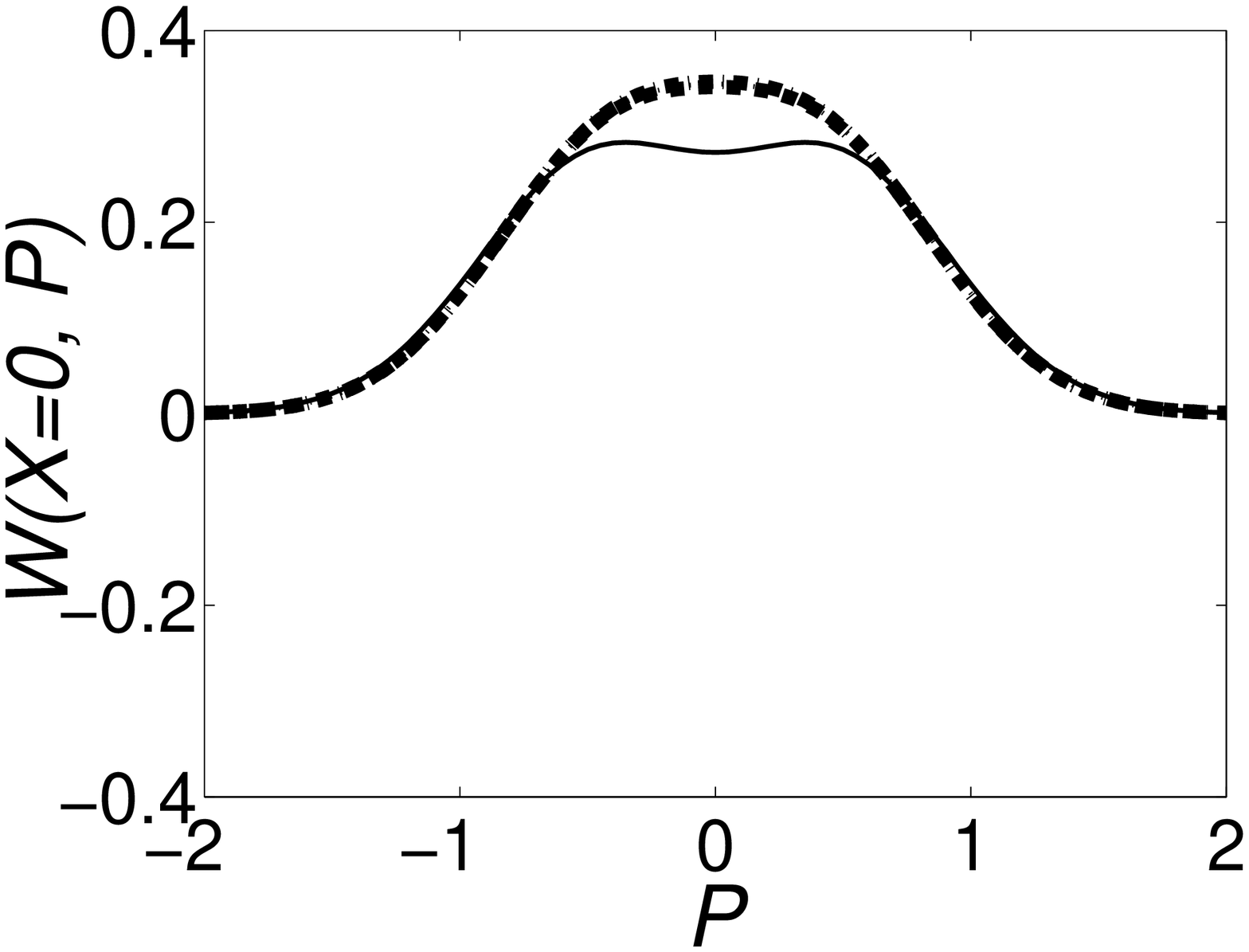}\hspace*{-0.5mm}\epsfxsize=3.6cm
\epsfbox{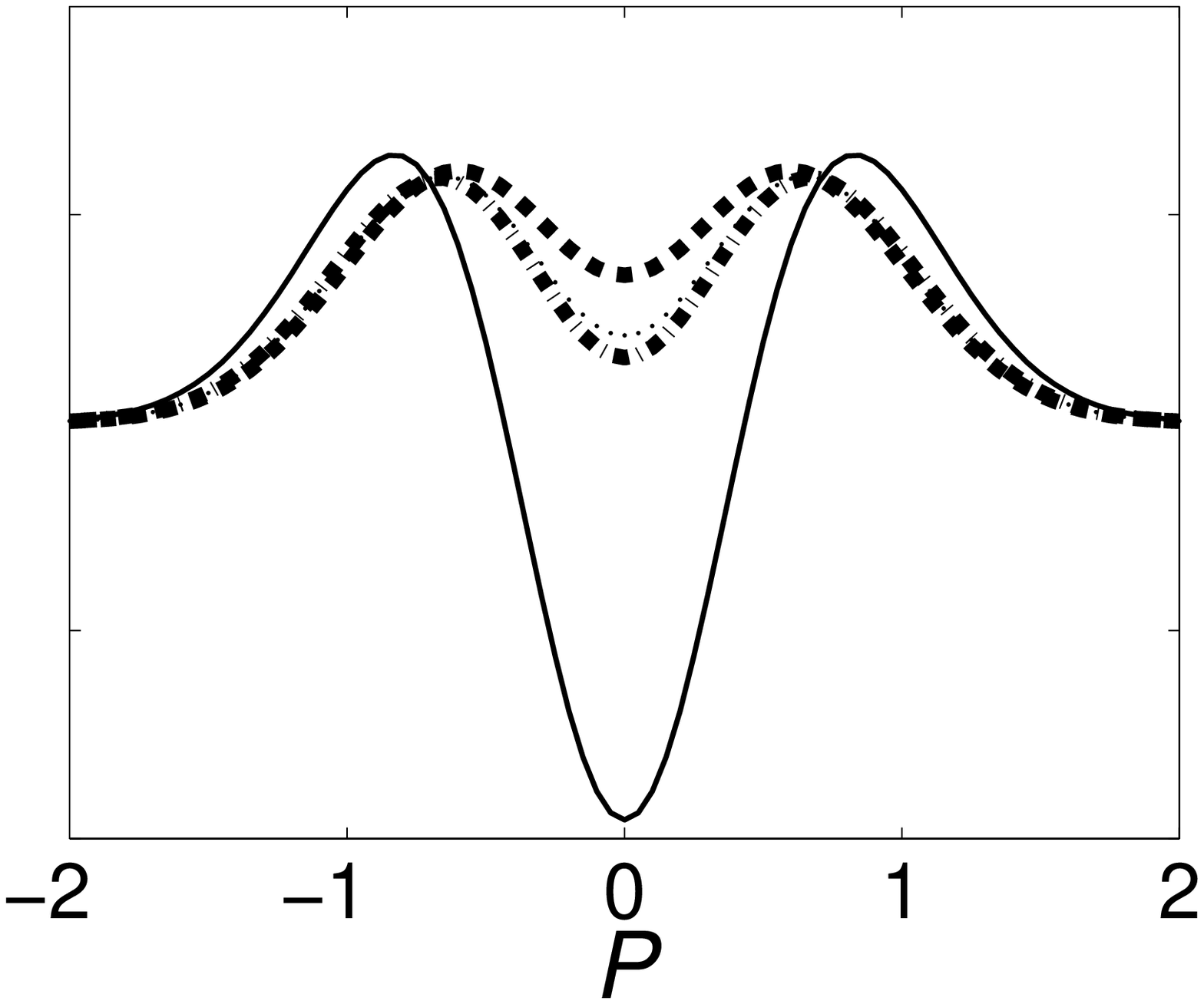}\vspace*{3mm} %
\caption{Effect of detector efficiency on Wigner function of the
output states generated by truncating coherent states of different
intensities: $|\alpha|^2=0.4$ (left) and $|\alpha|^2=4.0$ (right
figures). Solid curve is for the perfect QSD, and dash-dotted,
dotted, and dashed curves are for $\eta=0.5, 0.7$, and $1$,
respectively.}\label{fig11}
\end{figure*} \noindent %
of Wigner functions for different $|\alpha|^2$ and $\eta$ and
depicted some results in Figs. \ref{fig11}. We have understood
that for coherent input of low intensity $|\alpha|^{2}<1.0$,
detector losses and source imperfections do not have a significant
effect on the shape of the Wigner function. For $\eta\geq0.4$, the
shape of the Wigner function and marginal probabilities of the
states obtained from the proposed scheme and the desired state are
almost the same. For the range $1.0\leq|\alpha|^2 \leq1.5$,
negativity in $W(X,P=0)$ can be observed for $\eta\geq0.2$,
however, $W(X=0,P)$ becomes smoothed and the dip seen in the ideal
case fades with decreasing $\eta$. For $|\alpha|^{2}>1.5$, the
effect of detector efficiency is more profound. Moreover, it
affects $W(X,P=0)$ and $W(X=0,P)$ differently. For decreasing
$\eta$, although the value of negativity in $W(X,P=0)$ approaches
zero, the negativity can still be observed for $\eta>0.3$. On the
other hand, the dip
\linebreak%
\begin{figure*}[h]
\vspace*{-5mm}\hspace*{-5mm} \epsfxsize=4.5cm \epsfbox{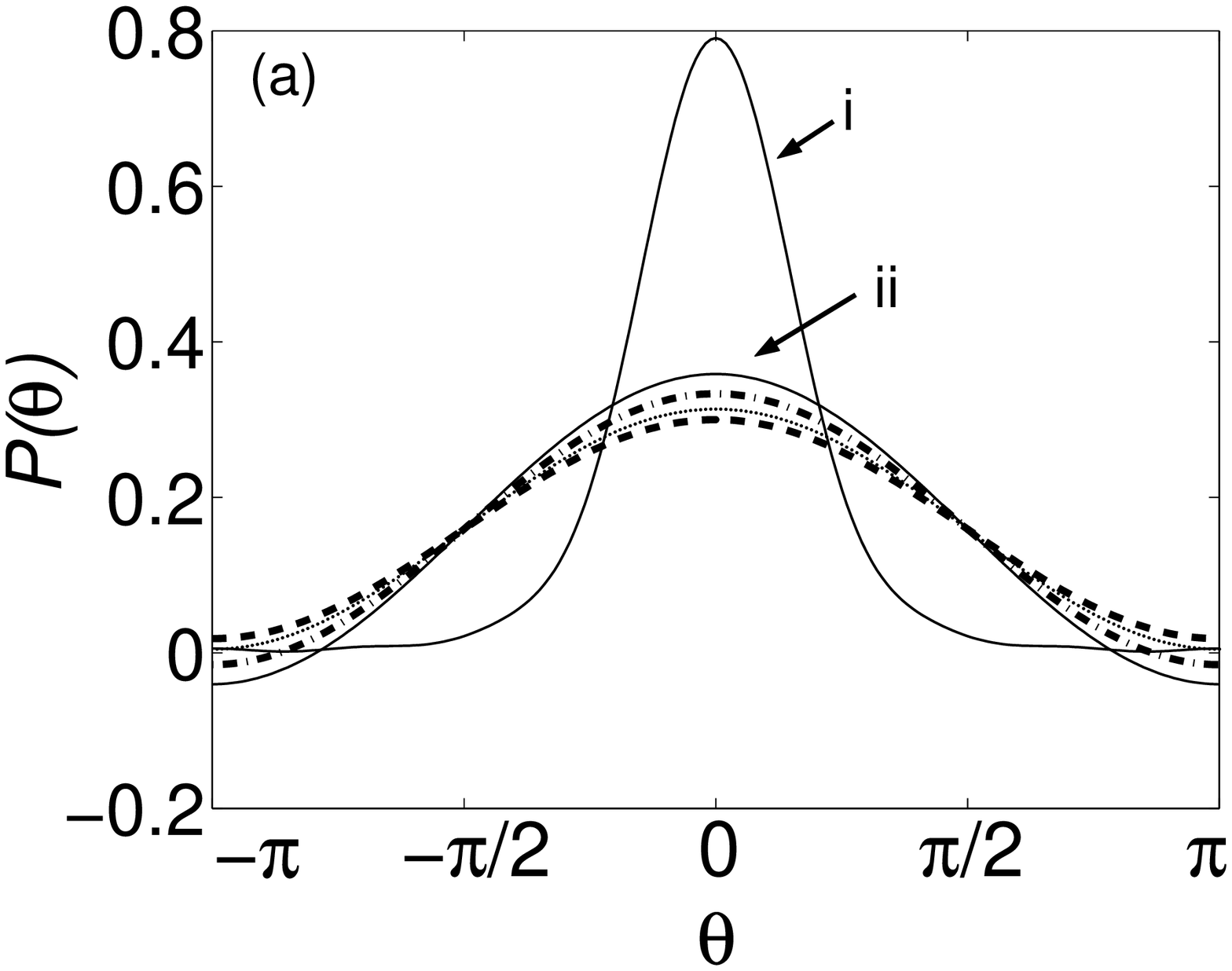}
\epsfxsize=3.8cm \epsfbox{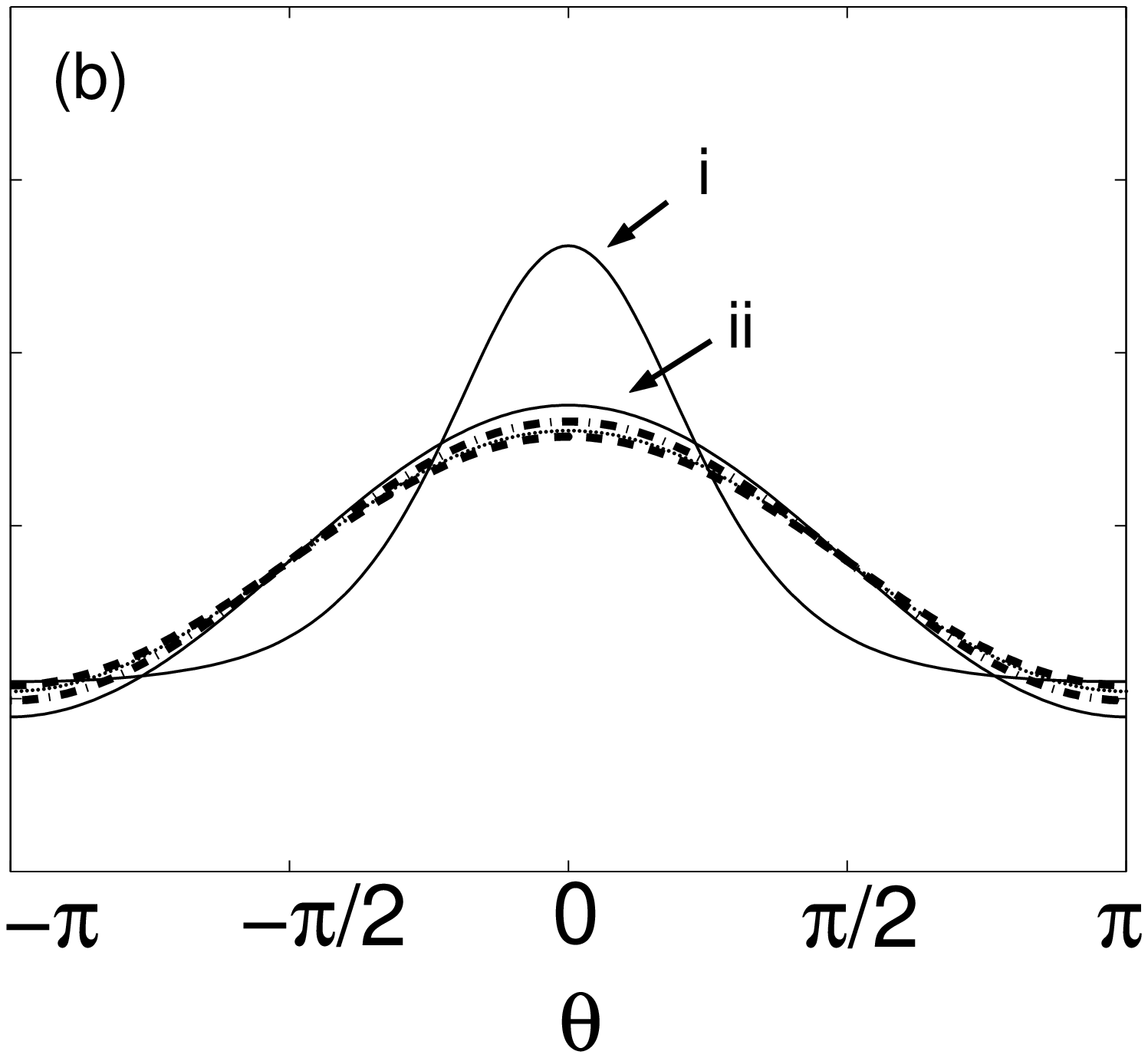} \vspace{2.5mm} %
\caption{Wigner phase distributions $P(\theta)$ of truncated
output states obtained from input coherent states of (a)
$|\alpha|^{2}=1.0$ and (b) $|\alpha|^{2}=0.4$. Solid curves
labeled as $(i)$ and $(ii)$ denote the phase distributions of the
input coherent states and the truncated output states with the
perfect QSD, respectively. Results of the proposed scheme are
shown as dash-dotted, dotted, and dashed curves for
$\eta=1.0,~0.5$, and $0.2$, respectively.} \label{fig12}
\end{figure*} \noindent %
seen in $W(X=0,P)$ is strongly smoothed monotonically with
increasing efficiency. Strong dips similar to the ideal case can
be observed for $0.2<\eta<0.5$. This deformation in the Wigner
function of the output state for high intensity input coherent
lights can be explained with the intrinsic property of the CPC's,
that is, they lose the information on the number of incident
photons. For strong lights, the number of photons in the system
will be higher than two photons, which will trigger ``false
alarms" about the generation of the desired state causing strong
deformations in the Wigner function.

Another important point of the QSD scheme is the preservation of
the relative phase between vacuum and single-photon components in
the ideal case, so it is necessary to study the phase and its
distribution for the output state. The effect of imperfections on
the phase distribution of the generated state is analyzed using
Wigner phase distribution, which is the phase distribution
associated with the Wigner function and calculated using
\cite{Tan96}
\begin{equation}
P(\theta)=\int_{0}^{\infty}W(\beta)|\beta| d|\beta|\, , \label{N25}
\end{equation}
where $W(\beta)$ can be obtained from Eqs.
(\ref{N23})--(\ref{N24}) using $\beta=X+\I P$.

Analysis of Wigner phase distribution for different intensities of
the input coherent light and different detector efficiencies has
revealed the following.

(i)~Maximum value of the phase distribution is obtained at the
phase of the input coherent light, which implies that the
preferred phase for the output state obtained from the
experimental scheme is the phase of the input light. It must be
noted that in the proposed model we have not included
phase-dependent losses.

(ii)~Low detector efficiency and losses in the system smooth and
broaden the phase distribution. In the limit $\eta=0$, phase
distribution is flat and $P(\theta_1)\sim0.1592$, which are also
observed for the vacuum state.

(iii)~For $0.25 < |\alpha|^{2} <0.45$, phase distribution has
negativity for the ideal scheme. To preserve the negativity of
phase distribution in the experimental scheme, detector efficiency
must be high, i.e., for $|\alpha|^{2}=0.4, 1.0$, and $2.0$,
detector efficiency $\eta$ must be greater than $0.9, 0.6$, and
$0.7$, respectively. However, the minimum value of $P(\theta)$ is
much higher than that of the ideal case even for $\eta=1$.

(iv)~Negativity of the Wigner phase distribution can be preserved
with $\eta\geq0.65$ when the weights of the vacuum and
single-photon states are comparable (nearly equal), because the
minimum of the phase distribution is more strongly negative for
these cases, i.e., in the ideal case, $|\alpha|^2=1$ and the
unnormalized state $|0\rangle+\alpha|1\rangle$, $P(\theta_1)$
becomes $\sim-0.0403$, which is the lowest minimum for any
$|\alpha|^2$ for this superposition state. Figure \ref{fig12}
shows the effect of detector efficiency on the Wigner phase
distribution for input states of different intensities.

\section{CONCLUSIONS}

In this study, we have analyzed the QSD scheme proposed by Pegg et
al. in detail using realistic descriptions for the detectors and
single-photon source. We have also proposed and discussed a simple
and realizable experimental scheme for the QSD considering
possible error sources that can be encountered in practice. The
possible ways of solving these difficulties and their effect on
the generated output state are discussed. We have shown explicitly
that the proposed scheme is realizable with high fidelity using
commercially available detectors and SPDC as the single-photon
source. With a further analysis using quasidistributions, it has
been clarified that the states prepared with the QSD scheme have
nonclassical properties, and one can distinguish these states from
states generated by other strategies.  Moreover, it has been shown
that the value of the preferred relative phase between the vacuum
and single-photon states of the input coherent light is preserved
at the truncated output state. The verification of truncation
process can be done using a combination of photon counting and
homodyning or by more sophisticated methods such as
homodyne-tomography techniques \cite{Welsch99}.

\acknowledgements

We thank Stephen M. Barnett, Takashi Yamamoto and Yu-Xi Liu for
stimulating discussions. This work was supported by a Grant-in-Aid
for Encouragement of Young Scientists (Grant No.~12740243) and a
Grant-in-Aid for Scientific Research (B) (Grant No.~12440111) by
Japan Society for the Promotion of Science.


{\setlength{\fboxsep}{10pt}
\begin{center}
\framebox{\parbox{0.75\columnwidth}{%
\begin{center}
to appear in\\ Phys. Rev. A {\bf 64}, 0638xx (2001)
\end{center}}}
\end{center}

\end{document}